\newif\ifplainstyle
\plainstyletrue
\plainstylefalse

\newif\ifjhepstyle
\jhepstyletrue

\newif\ifprstyle
\prstyletrue
\prstylefalse

\ifprstyle
	\documentclass[twocolumn,nofootinbib]{revtex4-1}
\else
	\documentclass[11pt,a4paper]{article}
\fi

\ifjhepstyle
	\usepackage{jheppub}
	\usepackage{amsfonts}
	\usepackage{verbatim}
	\usepackage{float}
	\usepackage{color}
	
	\usepackage{array}
	\newcolumntype{C}[1]{>{\centering\arraybackslash$}p{#1}<{$}}
	\makeatletter
	\def\@fpheader{\phantom{Prepared for submission to JHEP}}
	\makeatother
\else	
 	\ifprstyle
		\usepackage{verbatim}
		\usepackage{amsmath,amsfonts,amssymb}
		\usepackage[colorlinks=true
                	,urlcolor=blue
                	,anchorcolor=blue
                	,citecolor=blue
                	,filecolor=blue
                	,linkcolor=blue
                	,menucolor=blue
                	]{hyperref}
	\else
            	\usepackage{verbatim}
            	\usepackage{cite}
            	\usepackage{setspace}
            	\usepackage[top=2.5cm, bottom=2.75cm, left=2.5cm, right=2.5cm]{geometry}
            	\usepackage{amsmath,amsfonts,amssymb}
            	\usepackage[colorlinks=true
            	,urlcolor=blue
            	,anchorcolor=blue
            	,citecolor=blue
            	,filecolor=blue
            	,linkcolor=blue
            	,menucolor=blue
            	,linktoc=page
            	]{hyperref}
            	\usepackage{float}
            	\restylefloat{table}
            	
            	\numberwithin{equation}{section}
	\fi
\fi

\allowdisplaybreaks

\usepackage{tikz}
\usepackage{mathtools}
\usepackage{pgfplots}
\usepackage{subcaption}
\usepackage{bbold}
\usepackage{transparent}
\usepackage{bm}
\usepackage{enumitem}

\makeatletter
\let\save@mathaccent\mathaccent
\newcommand*\if@single[3]{%
  \setbox0\hbox{${\mathaccent"0362{#1}}^H$}%
  \setbox2\hbox{${\mathaccent"0362{\kern0pt#1}}^H$}%
  \ifdim\ht0=\ht2 #3\else #2\fi
  }
\newcommand*\rel@kern[1]{\kern#1\dimexpr\macc@kerna}
\newcommand*\widebar[1]{\@ifnextchar^{{\wide@bar{#1}{0}}}{\wide@bar{#1}{1}}}
\newcommand*\wide@bar[2]{\if@single{#1}{\wide@bar@{#1}{#2}{1}}{\wide@bar@{#1}{#2}{2}}}
\newcommand*\wide@bar@[3]{%
  \begingroup
  \def\mathaccent##1##2{%
    \let\mathaccent\save@mathaccent
    \if#32 \let\macc@nucleus\first@char \fi
    \setbox\z@\hbox{$\macc@style{\macc@nucleus}_{}$}%
    \setbox\tw@\hbox{$\macc@style{\macc@nucleus}{}_{}$}%
    \dimen@\wd\tw@
    \advance\dimen@-\wd\z@
    \divide\dimen@ 3
    \@tempdima\wd\tw@
    \advance\@tempdima-\scriptspace
    \divide\@tempdima 10
    \advance\dimen@-\@tempdima
    \ifdim\dimen@>\z@ \dimen@0pt\fi
    \rel@kern{0.6}\kern-\dimen@
    \if#31
      \overline{\rel@kern{-0.6}\kern\dimen@\macc@nucleus\rel@kern{0.4}\kern\dimen@}%
      \advance\dimen@0.4\dimexpr\macc@kerna
      \let\final@kern#2%
      \ifdim\dimen@<\z@ \let\final@kern1\fi
      \if\final@kern1 \kern-\dimen@\fi
    \else
      \overline{\rel@kern{-0.6}\kern\dimen@#1}%
    \fi
  }%
  \macc@depth\@ne
  \let\math@bgroup\@empty \let\math@egroup\macc@set@skewchar
  \mathsurround\z@ \frozen@everymath{\mathgroup\macc@group\relax}%
  \macc@set@skewchar\relax
  \let\mathaccentV\macc@nested@a
  \if#31
    \macc@nested@a\relax111{#1}%
  \else
    \def\gobble@till@marker##1\endmarker{}%
    \futurelet\first@char\gobble@till@marker#1\endmarker
    \ifcat\noexpand\first@char A\else
      \def\first@char{}%
    \fi
    \macc@nested@a\relax111{\first@char}%
  \fi
  \endgroup
}
\makeatother

\newcommand{\ThisIsTheTitle}{Exotic RG Flow of Entanglement Entropy} 
\newcommand{\ThisIsAuthorOne}{Chanyong Park}
\newcommand{\ThisIsEmailOne}{cyong21@gist.ac.kr}
\newcommand{\ThisIsAuthorTwo}{and Jung Hun Lee}
\newcommand{\ThisIsEmailTwo}{junghun.lee@gist.ac.kr}
\newcommand{\ThisIsTheAffiliation}{Department of Physics and Photon Science, Gwangju Institute of Science and Technology,
Gwangju, 61005, Korea}
\newcommand{\TheseAreTheKeywords}{}

\newcommand{\ThisIsTheAbstract}{In this paper, we holographically study the renormalization group (RG) flow in a three-dimensional Einstein-dilaton gravity with a potential permitting several types of the RG flow with nontrivial beta-functions. By using the intrinsic parameter of the potential, we classify possible holographic RG flows and examine their physical features. Using the Ryu-Takayanagi formulation, furthermore, we investigate how the $c$-function of the entanglement entropy behaves along the RG flow numerically. We show that the entanglement $c$-function monotonically decreases even in the cases with a nontrivial beta-function. For checking the consistency, we also compare the result of the entanglement entropy with the $c$-function derived from the holographic renormalization procedure.}

\ifjhepstyle
\title{\ThisIsTheTitle}

\author{\ThisIsAuthorOne}
\author{\ThisIsAuthorTwo}

\affiliation{\ThisIsTheAffiliation}

\emailAdd{\ThisIsEmailOne}
\emailAdd{\ThisIsEmailTwo}

\abstract{\ThisIsTheAbstract} 

\fi

\begin{document}


\ifjhepstyle
\maketitle
\flushbottom
\fi

\long\def\symfootnote[#1]#2{\begingroup%
\def\thefootnote{\fnsymbol{footnote}}\footnote[#1]{#2}\endgroup} 

\def\rednote#1{{\color{red} #1}}
\def\bluenote#1{{\color{blue} #1}}

\def\({\left (}
\def\){\right )}
\def\lb{\left [}
\def\rb{\right ]}
\def\lB{\left \{}
\def\rB{\right \}}

\def\Int#1#2{\int \textrm{d}^{#1} x \sqrt{|#2|}}
\def\Bra#1{\left\langle#1\right|} 
\def\Ket#1{\left|#1\right\rangle}
\def\BraKet#1#2{\left\langle#1|#2\right\rangle} 
\def\Vev#1{\left\langle#1\right\rangle}
\def\Vevm#1{\left\langle \Phi |#1| \Phi \right\rangle}\def\bbox{\bar{\Box}}
\def\til#1{\tilde{#1}}
\def\wtil#1{\widetilde{#1}}
\def\ph#1{\phantom{#1}}

\def\ra{\rightarrow}
\def\la{\leftarrow}
\def\lra{\leftrightarrow}
\def\p{\partial}
\def\diff{\mathrm{d}}

\def\sinh{\mathrm{sinh}}
\def\cosh{\mathrm{cosh}}
\def\tanh{\mathrm{tanh}}
\def\coth{\mathrm{coth}}
\def\sech{\mathrm{sech}}
\def\csch{\mathrm{csch}}

\def\pa{\partial}

\def\bl{\biggl}
\def\br{\biggr}
\def\a{\alpha}
\def\b{\beta}
\def\g{\gamma}
\def\d{\delta}
\def\e{\epsilon}
\def\ve{\varepsilon}
\def\k{\kappa}
\def\l{\lambda}
\def\n{\nabla}
\def\om{\omega}
\def\s{\sigma}
\def\t{\theta}
\def\z{\zeta}
\def\vp{\varphi}

\def\ss{\Sigma}
\def\dd{\Delta}
\def\GG{\Gamma}
\def\ll{\Lambda}
\def\tt{\Theta}

\def\A{{\cal A}}
\def\B{{\cal B}}
\def\C{{\cal C}}
\def\cE{{\cal E}}
\def\D{{\cal D}}
\def\F{{\cal F}}
\def\H{{\cal H}}
\def\I{{\cal I}}
\def\J{{\cal J}}
\def\K{{\cal K}}
\def\L{{\cal L}}
\def\N{{\cal N}}
\def\O{{\cal O}}
\def\P{{\cal P}}
\def\cS{{\cal S}}
\def\W{{\cal W}}
\def\X{{\cal X}}
\def\Z{{\cal Z}}

\def\mfa{\mathfrak{a}}
\def\mfb{\mathfrak{b}}
\def\mfc{\mathfrak{c}}
\def\mfd{\mathfrak{d}}

\def\we{\wedge}
\def\re{\textrm{Re}}

\def\tilw{\tilde{w}}
\def\tile{\tilde{e}}

\def\tilL{\tilde{L}}
\def\tilJ{\tilde{J}}

\def\zz{\bar z}
\def\xx{\bar x}
\def\yy{\bar y}
\def\xp{x^{+}}
\def\xm{x^{-}}

\def\bp{\bar{\p}}
\def\wei#1{{\color{red}#1}}

\def\VirU1{Vir \times U(1)}
\def\VirSL2R{\mathrm{Vir}\otimes\widehat{\mathrm{SL}}(2,\mathbb{R})}
\def\U1{U(1)}
\def\u1{U(1)}
\def\SL2R{\widehat{\mathrm{SL}}(2,\mathbb{R})}
\def\sl2r{\mathrm{SL}(2,\mathbb{R})}
\def\by{\mathrm{BY}}

\def\RR{\mathbb{R}}

\def\tr{\mathrm{Tr}}
\def\bnabla{\overline{\nabla}}

\def\sint{\int_{\ss}}
\def\dsint{\int_{\p\ss}}
\def\hint{\int_{H}}

\newcommand{\eq}[1]{\begin{align}#1\end{align}}
\newcommand{\eqst}[1]{\begin{align*}#1\end{align*}}
\newcommand{\eqsp}[1]{\begin{equation}\begin{split}#1\end{split}\end{equation}}
\def\bea{\begin{eqnarray}}
\def\eea{\end{eqnarray}}

\newcommand{\absq}[1]{{\textstyle\sqrt{\left |#1\right |}}}



\ifprstyle
\title{\ThisIsTheTitle}

\author{\ThisIsAuthorOne}
\email{\ThisIsEmailOne}

\author{\ThisIsAuthorTwo}
\email{\ThisIsEmailTwo}

\affiliation{\ThisIsTheAffiliation}

\date{\today}

\begin{abstract}
\ThisIsTheAbstract
\end{abstract}


\maketitle

\fi

\ifplainstyle
\begin{titlepage}
\begin{center}

\ph{.}

\vskip 4 cm

{\Large \bf \ThisIsTheTitle}

\vskip 1 cm

\renewcommand*{\thefootnote}{\fnsymbol{footnote}}

{{\ThisIsAuthorOne}\footnote{\ThisIsEmailOne} } and {\ThisIsAuthorTwo}\footnote{\ThisIsEmailTwo}}

\renewcommand*{\thefootnote}{\arabic{footnote}}

\setcounter{footnote}{0}

\vskip .75 cm

{\em \ThisIsTheAffiliation}

\end{center}

\vskip 1.25 cm

\begin{abstract}
\noindent \ThisIsTheAbstract
\end{abstract}

\end{titlepage}

\newpage

\fi

\ifplainstyle
\tableofcontents
\noindent\hrulefill
\bigskip
\fi


\section{Introduction} \label{intro}
Among contemporary theoretical physics, holography becomes one of the important research areas. Especially, after the AdS/CFT correspondence proposed in  \cite{Witten:1998qj,Maldacena:1999mh} this conjecture has provided us a new tool to understand quantum gravity and nonperturbative quantum features of the quantum field theory. Recently, the entanglement entropy has been paid attention in both string theory and condensed matter theory. The entanglement entropy measures the degrees of freedom confined in an arbitrarily chosen space-like subsystem. In two-dimensional conformal field theories, it has been shown that the coefficient of the universal term in the entanglement entropy is proportional to the central charge representing the degrees of freedom \cite{Calabrese:2004eu,Jafferis_2011,Cardy:1988cwa}. There are many attempts to generalize the two-dimensional result to the higher dimensional cases \cite{Myers_2011,Klebanov_2011}. Despite their salient property, the field theoretic computation accompanies with very complicated analysis. In this circumstance, the holographic calculation \cite{Ryu:2006bv,Ryu_2006} based on AdS/CFT correspondence \cite{Maldacena:1999mh,Gubser_1998,Witten:1998qj,Witten:1998zw} provides a more tractable tool because it enables us to interpret the entanglement entropy of strongly interacting systems as a geodesic of the dual classical Einstein gravity. In this work, we investigate the holographic renormalization group (RG) flow of the entanglement entropy and the property of the $c$-function along the RG trajectory when the boundary conformal field theory is deformed by a relevant operator \cite{Casini:2004bw,Casini:2006es,Albash:2011nq,Klebanov:2012yf,Cremonini:2013ipa,Faulkner:2014jva,Park:2014gja,Park:2015hcz,Casini:2015ffa,Kim:2016ayz,Bueno:2016rma,Kim:2016hig,Kim:2016jwu,Kim:2017lyx,Jang:2017gwd,Ghosh:2017big,Narayanan:2018ilr,Park:2018ebm}.

The AdS/CFT duality has provided the one-to-one map between non-perturbative conformal field theories (CFT)  and gravity/string theories defined in an one-dimensional higher AdS geometry at least in the large $N$ limit. 
Surprisingly,  it was shown that the holographic calculation of the entanglement entropy can reproduce exactly the same results obtained in a two-dimensional CFT \cite{Ryu:2006bv,Ryu_2006}. However, when we consider a CFT deformed by a relevant operator, the usual CFT description can not be used anymore at the low energy scale because the relevant deformation spoils the conformal symmetry. In other words, the relevant deformation can dramatically change the UV theory at the IR energy scale. Therefore, a CFT deformed by a relevant scalar operator gives us a chance to find a new CFT at an IR fixed point. In order to understand such a nontrivial RG flow from the holographic point of view, we consider an Einstein-dilaton gravity with an appropriate dilaton potential.  The gravity theory, which admits a smooth interpolation between UV and IR fixed points, has been known not only in gauged supergravity theories in $AdS_3$ but also other higher dimensional theories \cite{Deger_2000,Deger:2002hv}. Recently, the authors in \cite{Kiritsis:2016kog} found new solutions which show exotic behaviors of the RG flow. Related to such exotic RG solutions, in this work, we investigate how the dual field theory is modified along the RG flow by using the holographic renormalization and entanglement entropy techniques.  

One of the interesting and important tasks in physics may be counting the number of degrees of freedom of quantum field theories, which decreases monotonically along the RG flow. This feature called the $c$-theorem is well expressed by a $c$-function depending on the energy scale. For a suitably chosen entangling surface, the universal contribution to an entanglement entropy can be matched to the $c$-function \cite{Myers_2010,Myers_2011}. Such a $c$-function naturally reduces to the central charge at fixed points. Especially, in order to survey the diverse behaviors of $c$-functions for various RG flows, we introduce a specific dilaton potential with one free parameter, $a$. Relying on the value of $a$, in this work, 
we show that three different types of the RG flow, for instance, the standard, staircase and bouncing RG flows \cite{Kiritsis:2016kog}, are possible. For a positive value of $a$, the standard RG flow naturally appears. For a small negative value of $a$, the staircase RG flow appears and the $\beta$-function described by the dilaton field repeatedly changes the magnitude of its velocity without changing the sign along the RG trajectory. In a large negative value of $a$, interestingly, we obtain the bouncing RG solution which has a similar feature to the cascading RG flow studied in Ref. \cite{Kiritsis:2016kog}. In general, the cascading RG flow violates the Breitenlohner-Freedman (BF) bound at the UV energy scale. However, the bouncing solution we found does not violate the BF bound but changes the direction of the RG flow iteratively in the entire energy scale. Another intriguing point we observed is that in the bouncing RG solution, the number of the sign change in the $\beta$-function increases as the absolute value of $a$ becomes large.  We also check the $c$-functions of these flow solutions always decreases monotonically along the RG flow. 


The rest of this paper is organized as follows: In section \ref{sec2}, we review the basic setup for computing both a holographic RG flow and the corresponding $c$-function in the Einstein-dilaton gravity. To do so, we introduce the first order formalism where the superpotential plays a central role. In section \ref{sec3}, we take into account a toy model involving an appropriate potential with one free parameter. In a specific range of the parameter, the theory admits diverse geometric solutions interpolating two AdS spaces which, on the dual field theory side, represent various nontrivial RG flows from a UV to an IR fixed point. We classify the possible RG solutions and investigate their salient features. Applying the RT formulation, in section \ref{sec5}, we study the RG flow of the entanglement entropy with several different parameter values and investigate the corresponding $c$-theorem. Finally, we close this work with some concluding remarks in section \ref{discussion}.

\section{Holographic RG flow in Einstein-dilaton gravity} \label{sec2}
In this section, we investigate how an asymptotic AdS geometry deformed by a scalar field is connected to an RG flow and what kind of the RG flow can occur depending on the value of an intrinsic parameter. To do so, let us start with briefly reviewing the holographic RG flow.

Let us consider a Euclidean version of a minimally coupled Einstein-dilaton gravity with an arbitrary scalar potential
\bea			\label{act:EuclEins-scalar}
S =  - \frac{1}{2 \k^2} \int d^{d+1} X \sqrt{g} \left( {\cal R}  - \frac{1}{2} g^{MN}
\partial_M \phi \partial_N \phi - V(\phi) \right) + \frac{1}{\k^2} \int_{\partial {\cal M}} d^d x \sqrt{\g} \ K ,
\eea
where  $g_{MN}$ and $\g_{\mu\nu}$ indicate a bulk metric and an induced metric on the boundary respectively. Since the variation of the gravity action usually contains a radial derivative of the metric at the boundary, the Gibbons-Hawking term is usually required to get rid of such a radial derivative term. An extrinsic curvature, $K= g^{MN} K_{MN}$, is given by a covariant derivative of a unit normal vector. Assuming that the dilaton field depends only on the radial coordinate and that the boundary space is flat, the most general metric ansatz preserving the boundary's planar symmetry in the normal coordinate can be represented as
\bea		\label{met:normalcoord}
ds^2 = e^{2 A(y)} \delta_{\mu\nu} d x^{\mu} dx^{\nu}  + dy^2 ,
\eea
where the scale factor $e^{A(y)}$ measures the energy scale of dual field theory and the dilaton field $\phi(y)$ is interpreted as a running coupling of the RG flow. Here, the geometric solution is entirely determined by the scale factor $A(y)$. At a conformal fixed point where the geometry becomes an AdS space, the scale factor is simply $A(y)=-y/R$ where $R$ is the AdS curvature radius. In this description, we implicitly assumed that the asymptotic UV boundary is located at $y =  -\infty$, while the IR boundary appears at $y = \infty$. Hence, two fixed points of an RG flow are matched to two boundaries of the above metric ansatz \eqref{met:normalcoord}. The details of the geometry are governed by the equations of motion of $\phi(y)$ and $A(y)$
\bea
0 &=& 2 d (d-1) \dot{A}^2 - \dot{\phi}^2 + 2 V(\phi) , 	\label{eq:consteq}\\
0 &=& 2 (d-1) \ddot{A}   +   \dot{\phi}^2   ,  \label{eq:canmom} \\
0 &=& \ddot{\phi} + d \dot{A} \dot{\phi} - \frac{\partial V(\phi) }{\partial \phi}  ,  \label{eq:phdynamics}
\eea
where the dot indicates a derivative with respect to $y$. Above the first equation is a constraint and the second and third are  dynamical equations of $A$ and $\phi$. Note that only two of them are independent because combining the first and second equations automatically leads to the third one. As a consequence, (\ref{eq:consteq}) and (\ref{eq:canmom}) can be regarded as two independent equations determining the geometry up to boundary conditions. It is worth noting that $\dot{A}$ never increases because of $\ddot{A} \leq 0$ in \eqref{eq:canmom}. From the viewpoint of the holographic RG flow, a holographic $c$-function for $d=2$ is defined as 
\bea				\label{equation:cfunction}
c= - \frac{3}{2\,G\dot{A}}\,  , \label{hcfunction}
\eea
where the Newton constant is given by $G = 8 \pi \k^2$. Using this relation, the monotonically decreasing behavior of the $c$-function becomes manifest due to \eqref{eq:canmom}.

To understand the more details of the holographic RG flow, we introduce a superpotential \cite{Freedman:1999gp,deBoer:2000cz,Skenderis_2006,Papadimitriou_2011}
\bea		\label{ans:superpotential}
W(\phi) = - 2 (d-1) \dot{A}  .
\eea
Using this superpotential, the above dynamical differential equation can be reduced to 
\bea          \label{eq:independeq1}
  \dot{\phi}(y)=W'(\phi),
\eea 
where the prime indicates a derivative with respect to $\phi$. Moreover, the constraint equation reduces to
\bea		\label{eq:independeq2}
V = \frac{1}{2} \left( \frac{\partial W  }{\partial \phi} \right)^2 - \frac{d}{4(d-1)} W  ^2 .
\eea
As a result, two gravitational equations, (\ref{eq:consteq}) and (\ref{eq:canmom}), can be decomposed into three first-order differential equations, (\ref{ans:superpotential}), (\ref{eq:independeq1}) and (\ref{eq:independeq2}). If the boundary position moves from $y=-\infty$ to a finite distance of $y$, this change of the boundary position is associated with the change of the energy scale on the dual field theory side. Since the value of $\phi$ at the boundary is dual to the coupling constant of the dual field theory, (\ref{eq:independeq1}) represents the energy dependence of the coupling constant on the dual field theory side, which is related to a $\beta$-function  
\bea				\label{Relation:betaphi}
\beta(\phi)\equiv \frac{d \phi}{d A} = \frac{\dot{\phi}}{\dot{A}}
=-2(d-1)\frac{W'}{W} . \label{betafunction}
\eea
Noting that the superpotential is proportional to the inverse of the $c$-function ($W \sim 1/c$), we can easily check that the superpotential does not decreases along the RG flow
\bea
\frac{dW}{dy}\geq 0 , \label{holoctheorem}
\eea
which is another representation of the $c$-theorem.

\section{Various renormalization group flows} \label{sec3}

\begin{figure}
\centering
\includegraphics[width=0.55\textwidth]{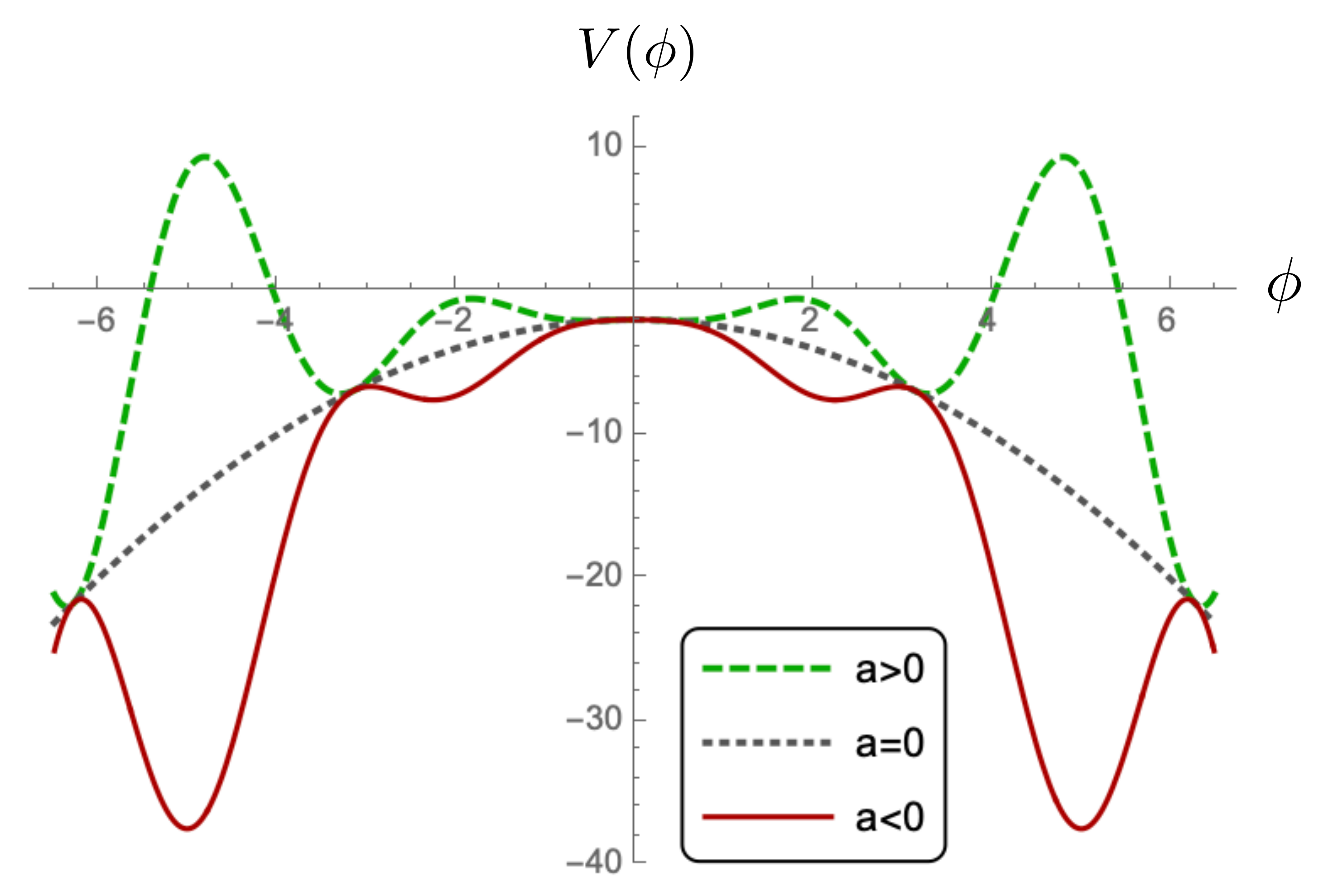}
\caption{ the potential depending the parameter $a$.}
\label{fig:pot}
\end{figure}

To study several different kinds of the RG  flow, we take into account a toy model having a specific scalar potential
\bea		\label{rel:potential}
V(\phi)=-\frac{d(d-1)}{R^2_{\text{uv}}}+ \frac{m^2}{2}  \phi^2 +a\phi^2\sin^2\phi. \label{potential}
\eea
where $m^2=- \Delta(d-\Delta)/R_{\text{uv}}^2$ and $a$ is a free parameter. In Fig. \ref{fig:pot}, we plot the potential with several different parameter values. Above, the first and second terms indicate a negative cosmological constant and a mass of the bulk scalar field, respectively. The last term describes a nontrivial interaction of the bulk scalar field. Notice that, when expanding the potential \eqref{rel:potential} around the UV fixed point $\phi_{\text{uv}_1}=0$, the potential gives rise to infinitely many extrema satisfying $V'(\phi_\ast)=0$. Since the potential \eqref{rel:potential} is invariant under $\phi\rightarrow-\phi$, the expansion of the potential is given by
\bea
V'(\phi)=(\phi-\phi_{\text{uv}_1})(\phi^2-\phi^2_{\text{ir}_1})(\phi^2-\phi^2_{\text{uv}_2})(\phi^2-\phi^2_{\text{ir}_2})\cdots, \label{vprime}
\eea   
where we assume $\phi_{\text{uv}_1}<\phi_{\text{ir}_1}<\phi_{\text{uv}_2}<\phi_{\text{ir}_2}<\cdots$. Hereafter, we focus only on the RG flow interpolating the first two fixed points, $\phi_{\text{uv}_1}$ and $\phi_{\text{ir}_1}$.

Assuming that $\phi$ approaches zero at the boundary, the asymptotic geometry near the boundary is given by a slightly deformed AdS geometry. In this case, $\phi$ and its derivative correspond to a coupling constant (or source) and a vacuum expectation value (vev) of the dual scalar operator in the dual field theory. Note that the asymptotic AdS geometry can appear only for $-d^2/4< m^2 R_{\text{uv}}^2 <0$. In this case, the dual scalar operator becomes a relevant operator. If $m^2$ is not in this parameter region, the dual scalar operator is irrelevant. On the gravity side, the corresponding bulk field $\phi$ diverges at the boundary and its gravitational backreaction modifies the asymptotic AdS geometry seriously. For the massless case with $m^2 =0$, the bulk scalar field is dual to a marginal operator \cite{Klebanov_1999,Witten:2001ua,Klebanov_2002}.  When we focus on the parameter region satisfying $-d^2/4< m^2 R_{\text{uv}}^2 <0$,  the leading behavior of the asymptotic AdS space is described by 
\bea
A(y) = - \frac{y}{R_{\text{uv}}} ,
\eea  
where the AdS radius $R_{\text{uv}}$ is associated with the degrees of freedom of the dual UV CFT. 

Note that in the asymptotic region, the mass of the bulk scalar field is associated with the conformal dimension, $\Delta$, of the dual scalar operator. The effect of the interaction term of the potential in \eqref{rel:potential} is negligible at least in the asymptotic region because we consider the relevant deformation. Even in this case, in the intermediate and IR regimes the effect of the interaction term becomes important and modifies the AdS geometry seriously into another geometry. In the asymptotic region, anyway, the gravitational backreaction of $\phi$ slightly modifies the asymptotic AdS geometry.  In this case, the profile of the scalar field is governed by the following equation of motion
\bea
0 = \ddot{\phi} - \frac{d}{R_{\text{uv}}} \dot{\phi} - m^2 \phi ,
\eea
and its solution  is given by
\bea
\phi =  c_1 \ e^{(d-\Delta) y/R_{\text{uv}}}    + c_2 \ e^{\Delta y/R_{\text{uv}}}  , \label{scalarsol}
\eea
where two integral constants, $c_1$ and $c_2$, are reinterpreted as a coupling constant (or source) and a vev of the dual scalar operator, as mentioned before.  Substituting the $\phi$ solution \eqref{scalarsol} into the equation of \eqref{eq:consteq}, the deformed geometry up to higher order corrections reduces to
\bea
A=- \frac{y}{R_{\text{uv}}} -\frac{c_1^2}{8(d-1)}\ e^{2(d-\Delta)y/R_{\text{uv}}}-\frac{c_2^2}{8(d-1)}\ e^{2\Delta y/R_{\text{uv}}}.
\eea

\begin{figure}
\centering
\begin{subfigure}{.45\textwidth}
  \centering
  \includegraphics[width=0.9\linewidth]{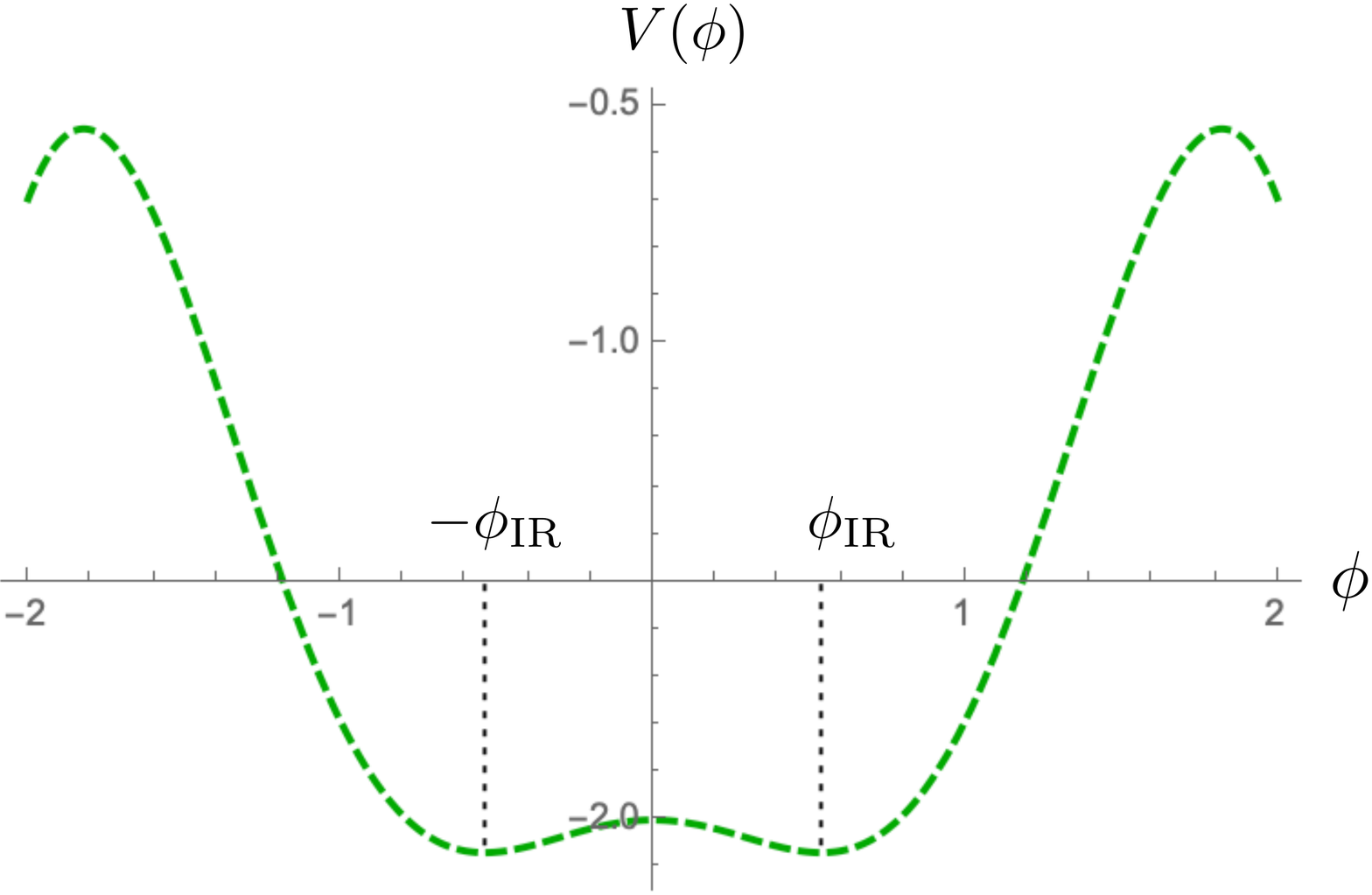}
  \caption{Potential with $a>0$.}
  \label{fig:pot1}
\end{subfigure}%
\begin{subfigure}{.45\textwidth}
  \centering
  \includegraphics[width=0.9\linewidth]{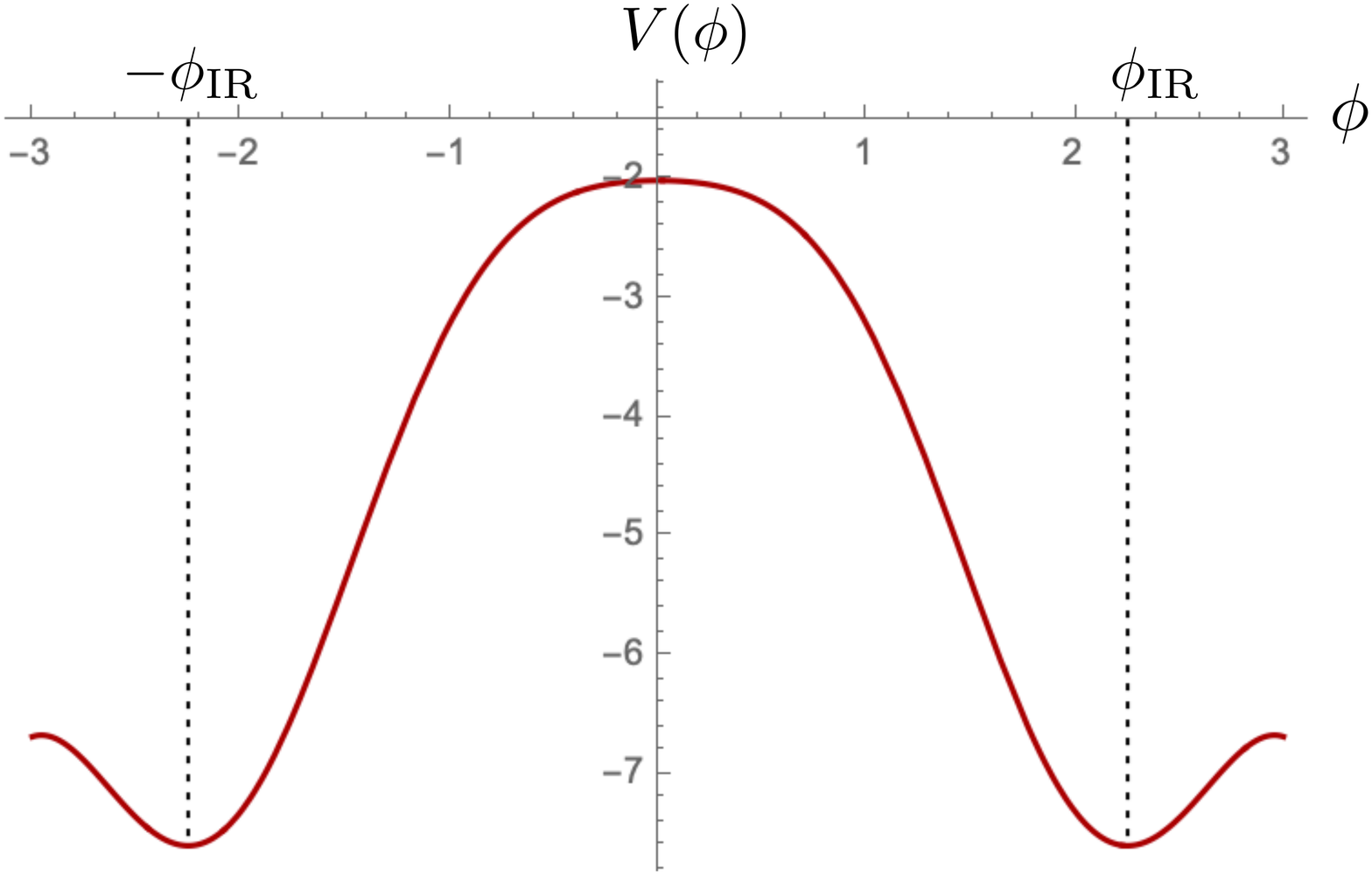}
  \caption{Potential with $a<0$.}
  \label{fig:pot2}
\end{subfigure}
\caption{Qualitative behavior of the potential  for $d=2$ and $\Delta=1$.}
\label{fig:pot12}
\end{figure}

\begin{figure}
\centering
\begin{subfigure}{.52\textwidth}
  \centering
  \includegraphics[width=1\linewidth]{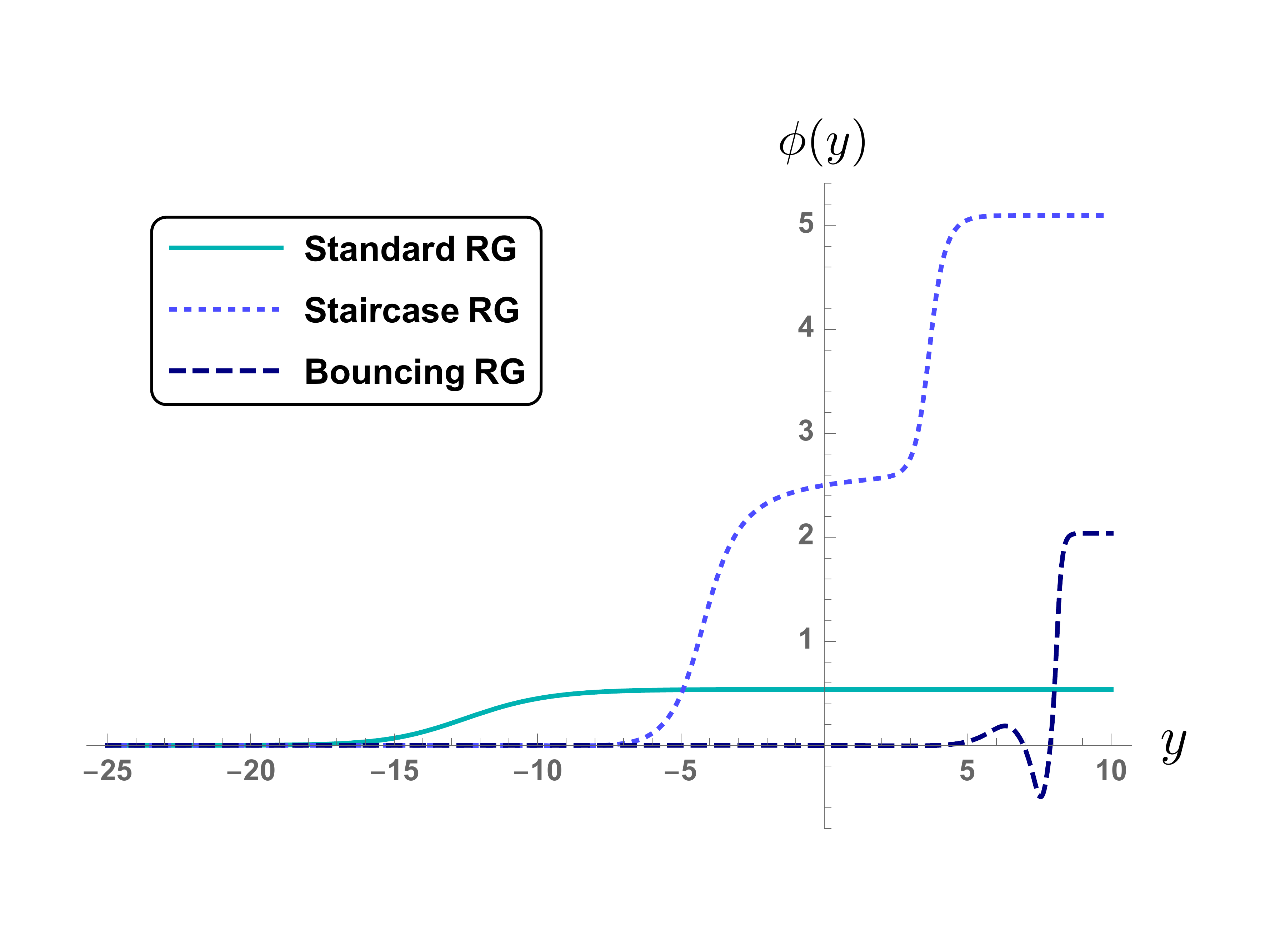}
  \caption{}
  \label{fig:SP}
\end{subfigure}%
\begin{subfigure}{.52\textwidth}
  \centering
  \includegraphics[width=1\linewidth]{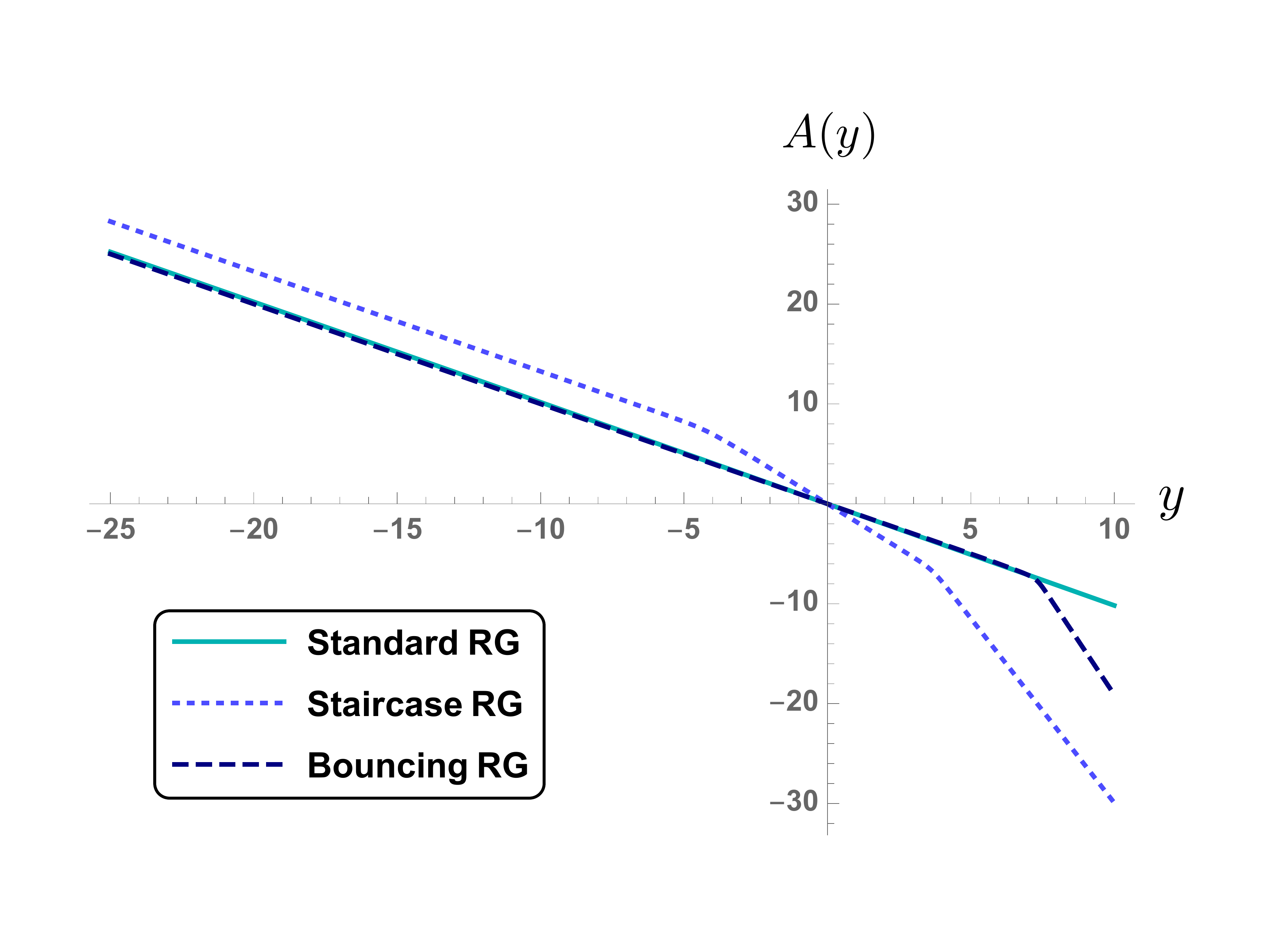}
  \caption{}
  \label{fig:SF}
\end{subfigure}
\caption{The profiles of (a) the scalar field $\phi(y)$ and (b) the metric factor $A$, which represent the standard ($a=1$), staircase $(a=-0.55)$ and bouncing $(a=-20)$ RG flows, respectively. }
\label{fig:SolProfiles}
\end{figure}

Near the UV fixed point with $\phi =0$, we depict the qualitative behavior of the potential relying on $a$ in Fig. \ref{fig:pot12}, which shows that the UV fixed point is unstable regardless of the value of $a$. This means that $\phi$ must roll down to another stable vacuum. Recalling that the value of $\phi$ can be reinterpreted as the coupling constant, the rolling of $\phi$ is related to the $\beta$-function discussed before. As shown in Fig. \ref{fig:pot12}, there exist two local minima near $\phi=0$ which correspond to the IR vacuum. The nearest two local minima are always degenerate due to the invariance of the potential under $\phi \to - \phi$. From now on, we concentrate on the case with  $\phi = \phi_{\text{ir}}$ and investigate the RG flow from the UV fixed point with $\phi=0$ to the IR fixed point with $\phi= \phi_{\text{ir}}$. The IR fixed point is stable and has a different vacuum energy $V(\phi_{\text{ir}})$ from $V(0)$ at the UV fixed point. On the dual field theory side, the different vacuum energy means that another conformal field theory appears with a different central charge at the IR fixed point. As a result, a geometric solution interpolating two extrema with the rolling $\phi$ describes the RG flow of the dual field theory from the UV to IR fixed points with a nontrivial $\beta$-function. In Fig. \ref{fig:SolProfiles}, we plot numerical solutions satisfying all equations of motion with several different parameter values. Relying on the value of $a$, these numerical solutions can show several different types of the RG flow including an exotic RG flow studied in Ref. \cite{Kiritsis:2016kog}. Now, we discuss more details about the possible RG flows.

\subsection{Standard and IR incomplete RG flows for $a \ge 0$}

Here, we briefly summarize the standard and IR incomplete RG flows which usually appear in many holographic models. \\

\noindent {\bf (1) Standard RG flow}  

For $a>0$, the local extremum at $\phi=0$ becomes a unstable local maximum.  In this case, a stable local minimum always exists near the local maximum. Therefore, the scalar field $\phi$ naturally rolls down from the local maximum to the other local minimum. On the dual field theory side, this behavior corresponds to the RG flow from the UV to IR fixed points. Moreover, the rolling of $\phi$ is related to the $\beta$-function of the dual field theory. Recalling that the $c$-function monotonically decreases with $c>0$ along the RG flow, we can easily see that $\dot{A}$ always has a negative value, $\dot{A} = -3/(2 G c) < 0$. This feature is also manifest in Fig. \ref{fig:SF}. Using this fact, the $\beta$-function in \eqref{betafunction} can be reexpressed by $\b \sim - c \dot{\phi}$. If the $\b$-function is not oscillating during the RG flow, we call it the standard RG flow. The solution for the standard RG flow was constructed by using a variety of potentials (see Ref. \cite{Park:2018ebm} and references therein for more details). \\

\noindent {\bf (2) Incomplete IR RG flow}

For $a=0$ unlike the case with $a>0$, there is no local minimum near the local maximum at $\phi=0$. In this case, the scalar field rolls down forever and finally diverges in the IR limit. This fact indicates that there is no stable IR fixed point which stops the rolling of the scalar field. From the dual field theory viewpoint, the absence of the local minimum corresponding to the IR fixed point implies the incompleteness of the dual field theory. As a result, the rolling of the scalar field for $a=0$ describes an IR incomplete RG flow of the dual field theory.

\subsection{Exotic (or novel) RG flow for $a < 0$}

For $a<0$, a new local minimum near the local maximum again appears. This fact, on the dual field theory side, indicates that there exists a new stable IR fixed point. For $a<0$, however, the RG flow can show a different behavior from the above standard RG flow, which was known as the exotic (or novel) RG flow \cite{Kiritsis:2016kog}. Relying on the behavior of the $\beta$-function, the exotic RG flow can be further classified to staircase and bouncing RG flows.   \\

\noindent {\bf (1) Staircase RG flow}  

\begin{figure}
 \centering
\begin{subfigure}{.45\textwidth}
  \centering
  \includegraphics[width=1\linewidth]{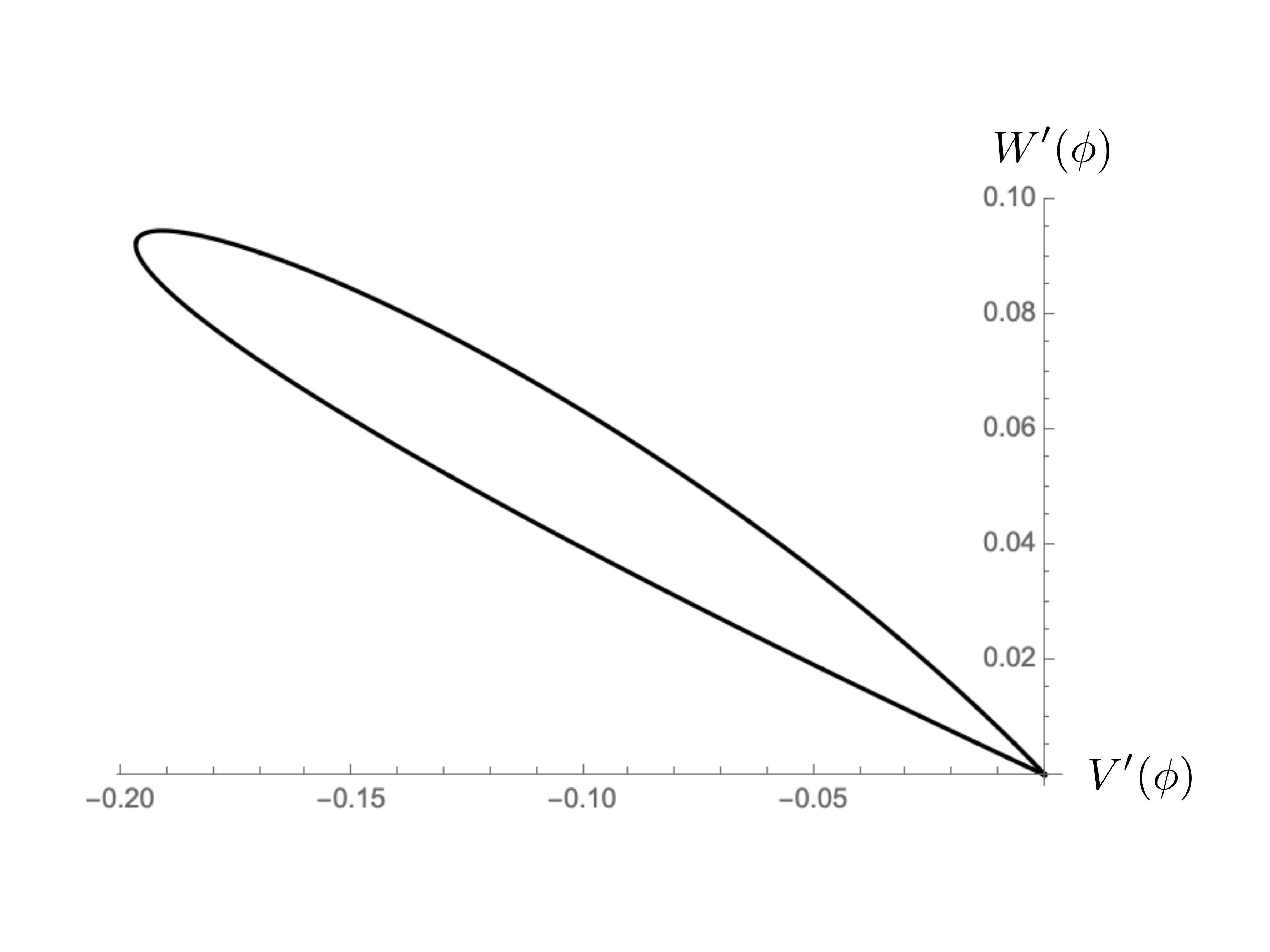}
  \caption{Phase curve in the standard RG flows}
  \label{fig:phaseS}
\end{subfigure}
\begin{subfigure}{.45\textwidth}
  \centering
  \includegraphics[width=1\linewidth]{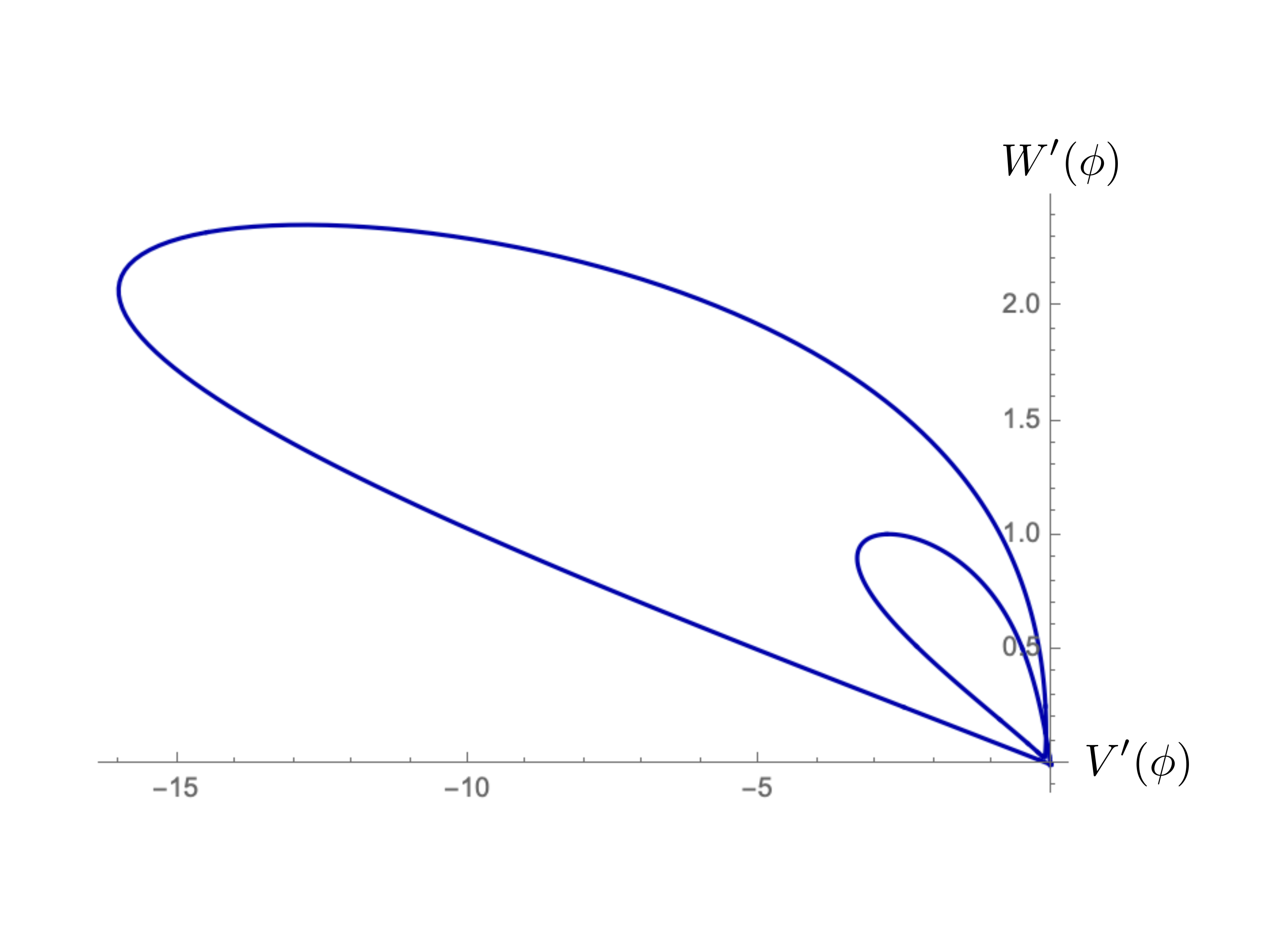}
  \caption{Phase curve in the staircase RG flows}
  \label{fig:phaseC}
\end{subfigure}
\begin{subfigure}{.44\textwidth}
  \centering
  \includegraphics[width=1\linewidth]{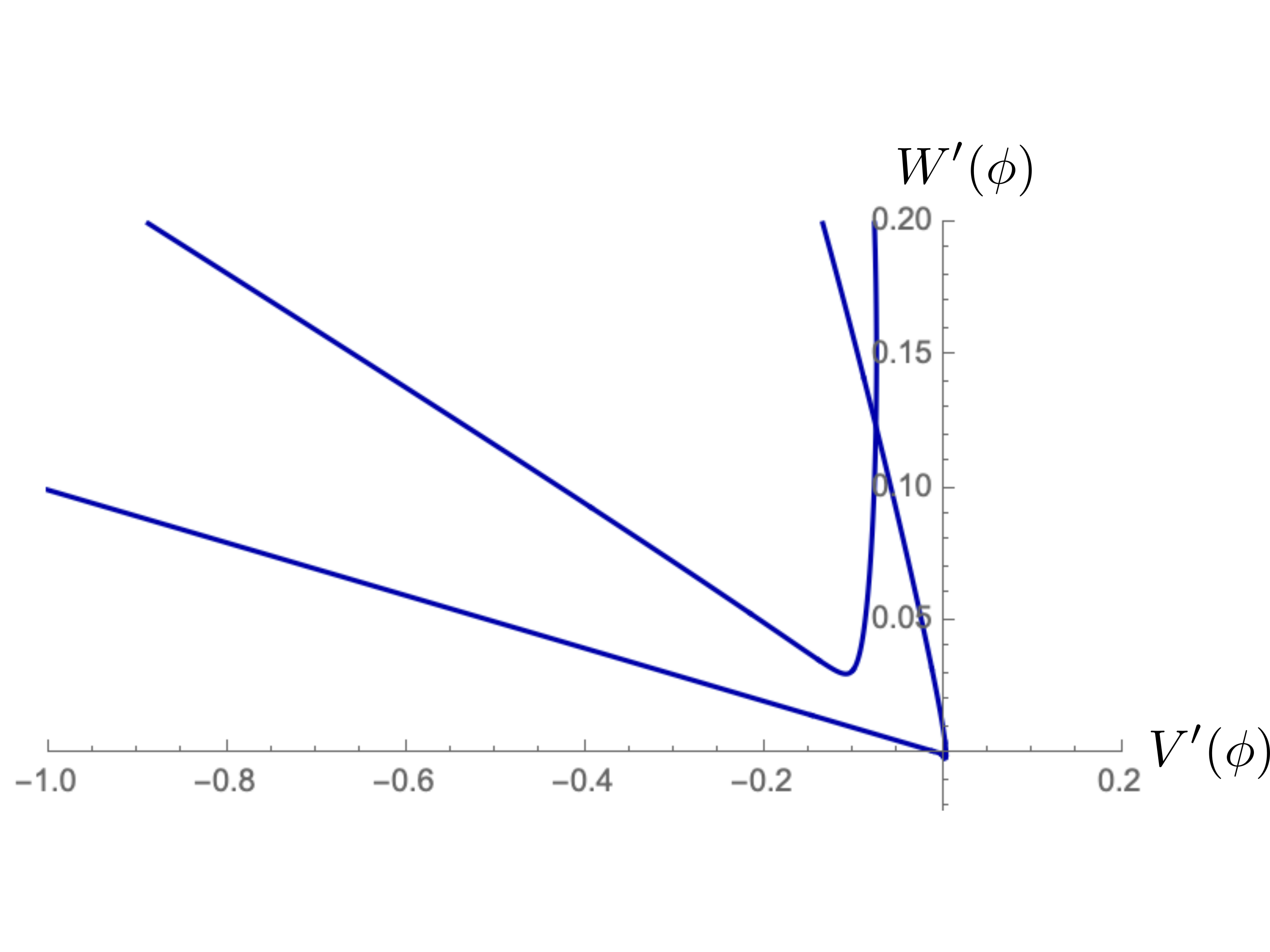}
  \caption{Phase curve near the origin of Fig. (b)}
  \label{fig:phaseCC}
\end{subfigure}
\caption{Phase diagram with $a=1$ (a) and $a=-0.55$ (b), respectively. The two curves (a) and (b) flow counterclockwise starting from the origin and terminate their flowing at the same origin. Notice that we distinguish the original point as the UV and the IR fixed points by mapping it into the field or holographic coordinate space. One can easily find that except for the origin there are no intersection points ($W'=0, V'\neq 0$) in the phase space.}
\label{fig:phaseSC}
\end{figure}

When $a$ has a small negative value ($a=-0.55$), the numerical solution in Fig. \ref{fig:SolProfiles} leads to a staircase RG flow which, at first glance, seems to have several plateaux satisfying $\dot{\phi}=0$. Recalling that  $\dot{\phi} \sim W'(\phi)$ is related to the $\beta$-function in \eqref{Relation:betaphi}, the existence of $\dot{\phi}=0$ implies that additional fixed points except the UV and IR fixed points can exist in the staircase RG flow. In Fig. \ref{fig:phaseSC}, we depict $W'(\phi)$ and $V'(\phi)$ for the standard and staircase RG flows. In the standard RG flow in Fig. \ref{fig:phaseS}, the UV and IR fixed points corresponding to the ends of the RG flow satisfies $W'(\phi) = V'(\phi) = 0$ simultaneously. In this case, $W'(\phi) = 0$ means that the $\beta$-function vanishes and $V'(\phi) = 0$ indicates that the corresponding dual theory is in a stable or unstable equilibrium state. These two conditions are natural requirements to obtain the geometry dual to the ground state of conformal field theories at UV and IR fixed points. In Fig. \ref{fig:phaseC}, however, the staircase RG flow, as mentioned before, looks to allow an additional fixed point satisfying $W'(\phi) = V'(\phi) = 0$ in the intermediate region of the RG flow. In order to check whether such an addition fixed point really exists or not, we zoom in the region around the origin of Fig. \ref{fig:phaseC} and plot the result in Fig. \ref{fig:phaseCC}. Fig. \ref{fig:phaseCC} shows that there is no point satisfying $W'(\phi) = V'(\phi) = 0$ exactly except the UV and IR fixed points.

\begin{figure}
\centering
\includegraphics[width=0.5\textwidth]{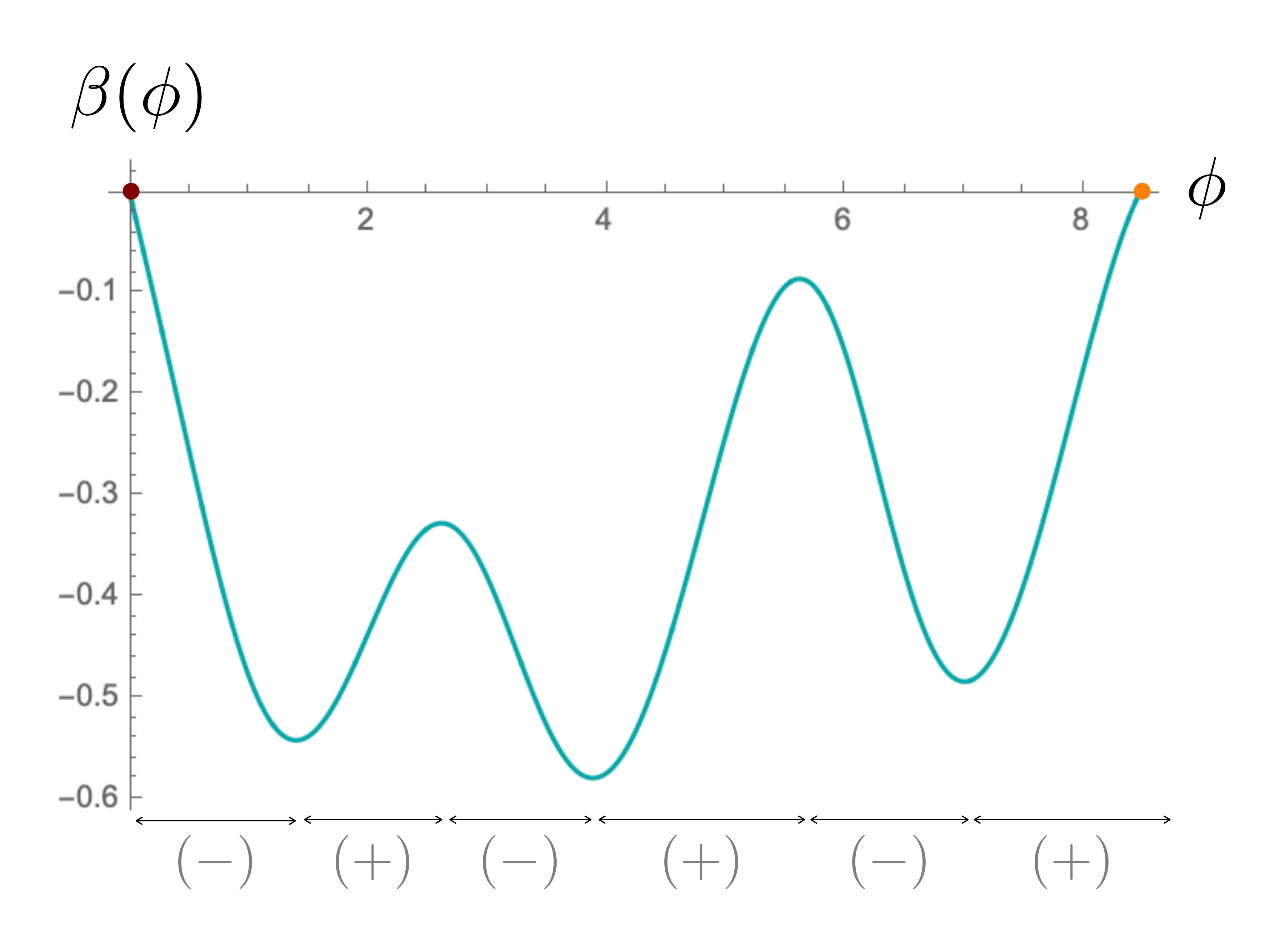}
\caption{The $\beta$-function of the staircase RG flow where two ends of the RG flow denote the UV and IR fixed points. }
\label{betaspeed}
\end{figure}

The absence of an additional fixed point discussed above becomes more manifest when we consider the $\beta$-function proportional to $W'$. We depict the $\beta$-function as a function of $\phi$ in Fig. \ref{betaspeed}, where there is no fixed point with a vanishing $\beta$-function except two UV and IR fixed points. In other words, the $\beta$-function does not change its sign during the RG flow. It is worth noting that the $\beta$-function appearing in Fig. \ref{betaspeed} always has a negative value except for the two fixed points. This fact means that the coupling constant of the dual field theory monotonically increases along the RG flow. Even in this case, since the magnitude of the $\beta$-function oscillates, the coupling constant repeats the fast and slow increasing during the RG flow. We called this type of the RG flow the staircase RG flow. \\

\noindent {\bf (2) Bouncing RG flow}

For a negatively large value of $a$ ($a=- 20$ in Fig. \ref{fig:SolProfiles}), the profile of $\phi$ in Fig. \ref{fig:SP} shows a totally different behavior from the above staircase RG flow. $\phi$ is monotonically increases in the staircase RG flow, whereas it oscillates in the case with a negatively large $a$. Since $\dot{\phi}$ is proportional to the $\beta$-function, the oscillation of $\phi$ indicates that there exists a point with a vanishing $\beta$-function at which the $\beta$-function changes the sign unlike the staircase RG flow. This type of the RG flow was known as the bouncing RG flow \cite{Kiritsis:2016kog}. 

To understand the more details of the bouncing RG flow, we introduce a new function by
\bea
B(\phi)=\sqrt{-\frac{4(d-1)}{d}V(\phi)}  . \label{Bcurve}
\eea
This new function is well defined only for $V \le 0$. In the present work, we consider the RG flows represented by $\phi$ with the range of $\phi_{uv} \le \phi \le \phi_{ir}$, so the value of $V$ during the RG flow always has a negative value satisfying $V(\phi_{ir}) < V(\phi_{uv}) < 0$. Therefore, the $B(\phi)$ we introduced is a well-defined positive function in the entire range of the RG flow. Rewriting $W' (\phi)$
in terms of the new positive function $B(\phi)$ allows two possible branches
\bea
W'(\phi)\equiv\frac{dW}{d\phi}=\pm\sqrt{\frac{2d}{d-1}(W^2-B^2)}.   \label{branchSol}
\eea

\begin{figure}
\centering
\begin{subfigure}{.47\textwidth}
  \centering
  \includegraphics[width=1\linewidth]{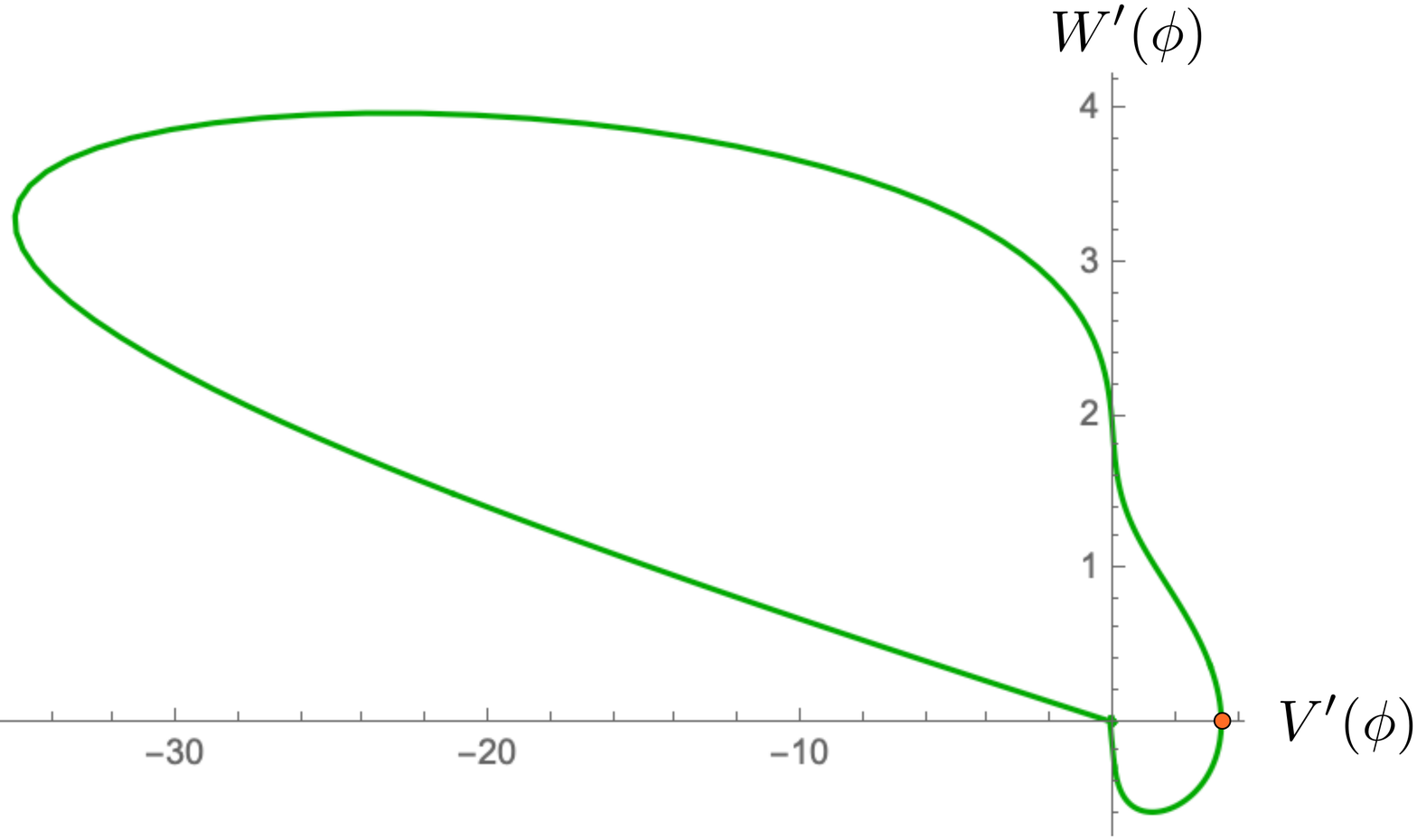}
  \caption{}
  \label{fig:phaseB1}
\end{subfigure}%
\begin{subfigure}{.47\textwidth}
  \centering
  \includegraphics[width=1\linewidth]{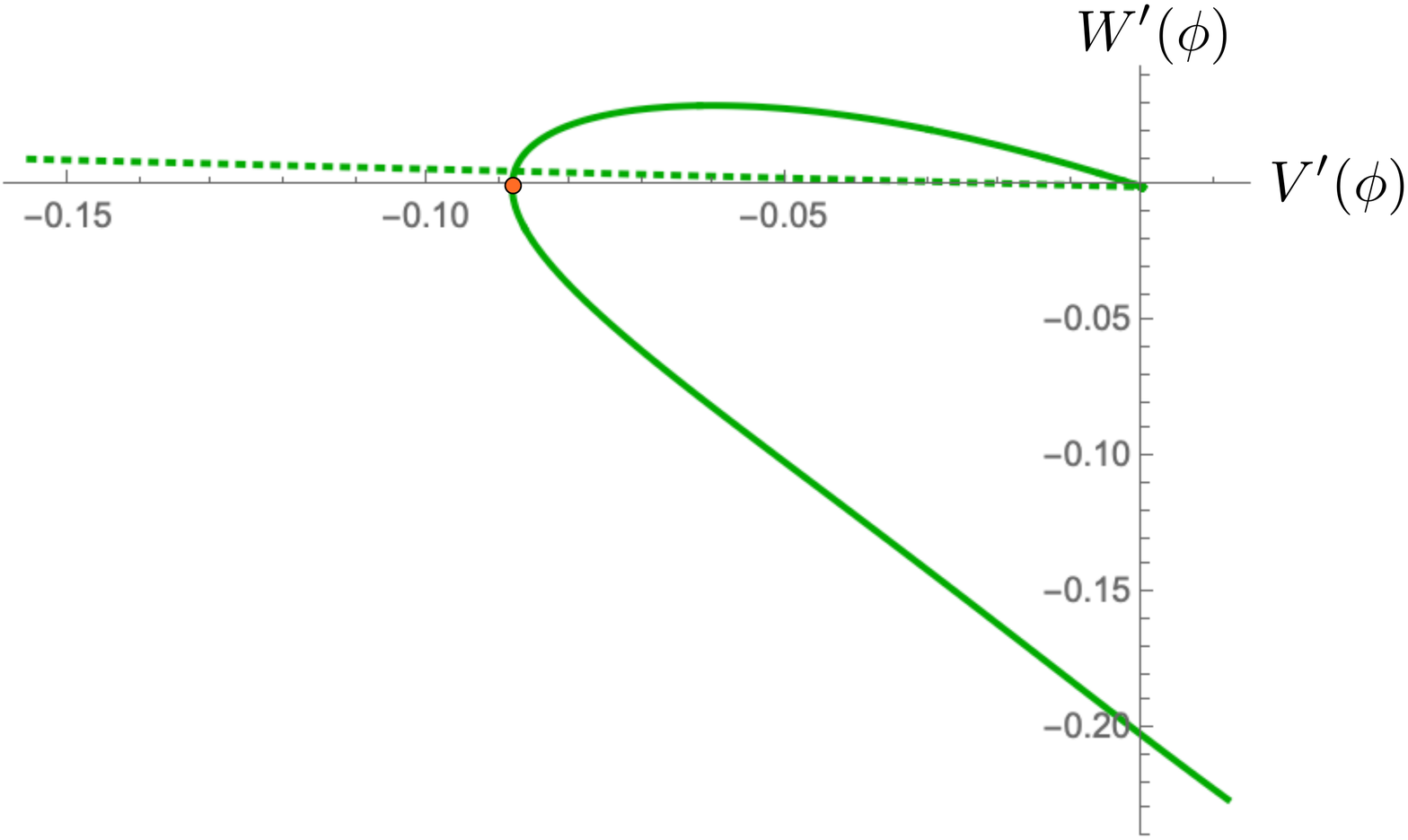}
  \caption{}
  \label{fig:phaseB2}
\end{subfigure}
\caption{Phase diagram with $a=-10$. The two curves (a) and (b) represent two bouncing points. The phase curve starts at the UV fixed point (the origin $W'=V'=0$) and flows counterclockwise and regresses into the origin (the IR fixed point). In the course of a flowing, the phase curve intersects horizontal line $W'=0$ twice (the orange circles). The function $W'$ at the intersection point changes its sign : $+\Rightarrow -$ in (b), $-\Rightarrow +$ in (a), which implies that at the critical $\phi_B$ ($W'(\phi_B)=0, V'(\phi_B)\neq 0$) the RG flow is bounced and inverts its direction. The dotted line curve in (b) represents an ingoing flow to the origin.}
\label{fig:phaseB}
\end{figure}

At this stage, there are several remarkable points we should note. First, the relation we obtained restricts the range of $W$ to the case of $W \ge B$ because the inside of the square root must be non-negative. Second, $W'$ vanishes at $W=B$ and two possible branches are smoothly connected at least up to the first derivative order. Suppose that the gravity solution has a point satisfying $W=B$ in the intermediate range of $y$. Then, the solution of one branch must smoothly change into the one of the other branch because the region satisfying $W^2<B^2$ is forbidden. This feature makes the RG flow bounce back at $W=B$ and leads to a vanishing $\beta$-function. Due to these reason, the RG flow showing this feature is called the bouncing RG flow. In Ref. \cite{Kiritsis:2016kog,Gursoy:2018umf,Ghosh:2017big}, the authors explored the exotic RG flows in different models and showed that bouncing solutions have vanishing $\beta$-functions unlike the standard and staircase RG flows, as explained before. Lastly, although $W' = 0$ at the bouncing point leads to a vanishing $\beta$-function, it does not guarantee $V'=0$. In general, the point with $W'=0$ is not coincident with the point satisfying $V'=0$ except for the UV and IR fixed points. In Fig. \ref{fig:phaseB}, we depict the curves of $W'$ and $V'$ for $a=-10$ where two bouncing points exist. As mentioned before, the bouncing points with $W'=0$ do not satisfy $V'=0$ in the intermediate region, $\phi_{uv} < \phi < \phi_{ir}$. In summary, the bouncing RG flow is the RG flow having the bouncing points, where the $\beta$-function vanishes, but it generally does not lead to additional conformal fixed points due to $V' \ne 0$. Then, what is the meaning of $\beta=0$ in the bouncing RG flow? A vanishing $\beta$-function in the bouncing RG flow is directly related to the change of the $\beta$-function's sign. This fact implies that the energy dependence of the coupling constant dramatically changes at the bouncing points. To understand this point more precisely, we need to remember that the positive or negative $\beta$-function usually means that the coupling constant decreases or increases along the RG flow, respectively. Therefore, the existence of a bouncing point at an intermediate energy scale indicates that a coupling constant increasing along the RG flow starts to decrease after passing through the bouncing point, or vice versa. In other words, the interaction strength of the bouncing RG flow does not monotonically increases along the RG flow.

\begin{figure}
 	\centering
 	\begin{subfigure}{.47\textwidth}
 		\centering
 		\includegraphics[width=1\linewidth]{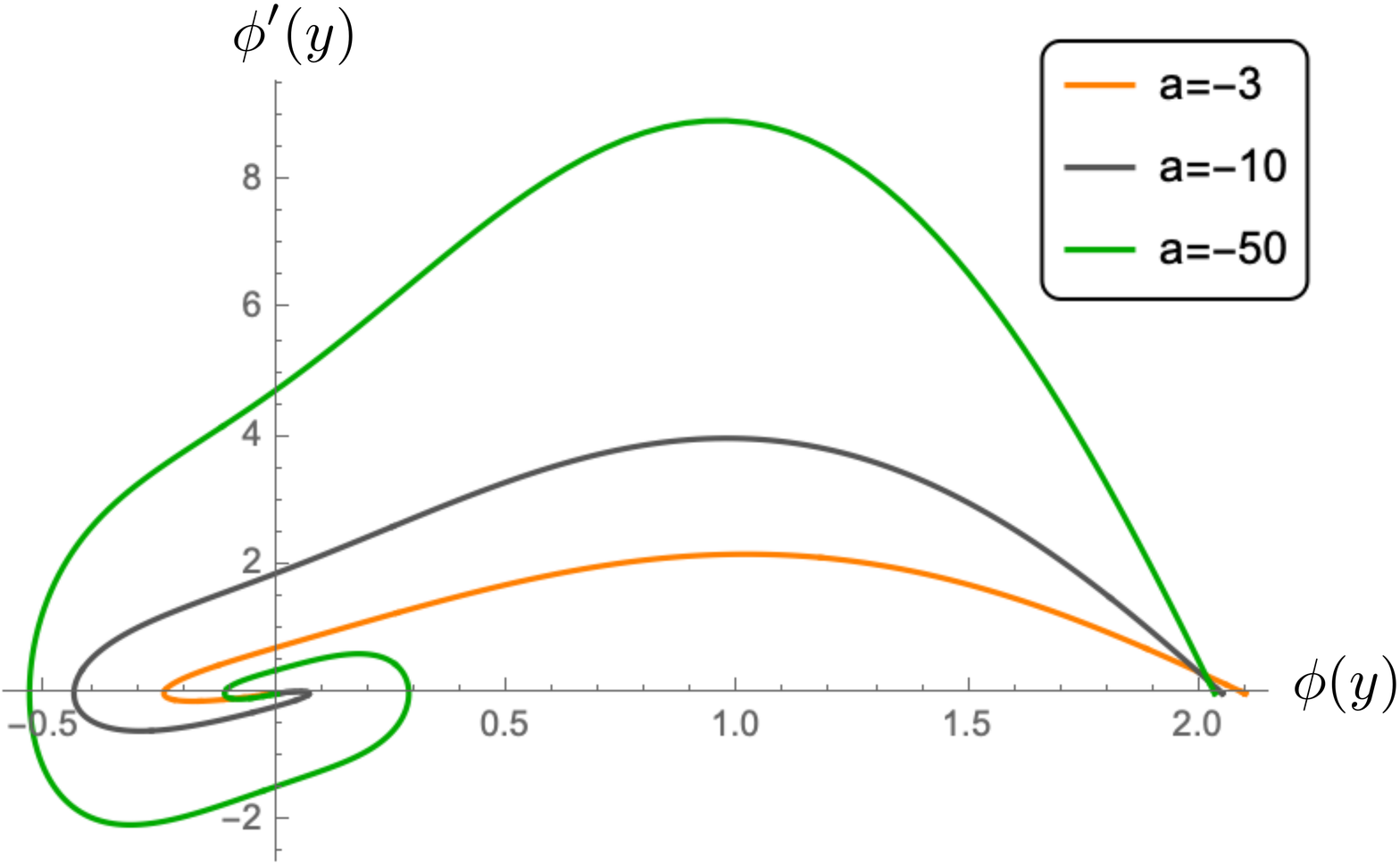}
 		\caption{}
 		\label{fig:phaseN1}
 	\end{subfigure}
 	\begin{subfigure}{.47\textwidth}
 		\centering
 		\includegraphics[width=1\linewidth]{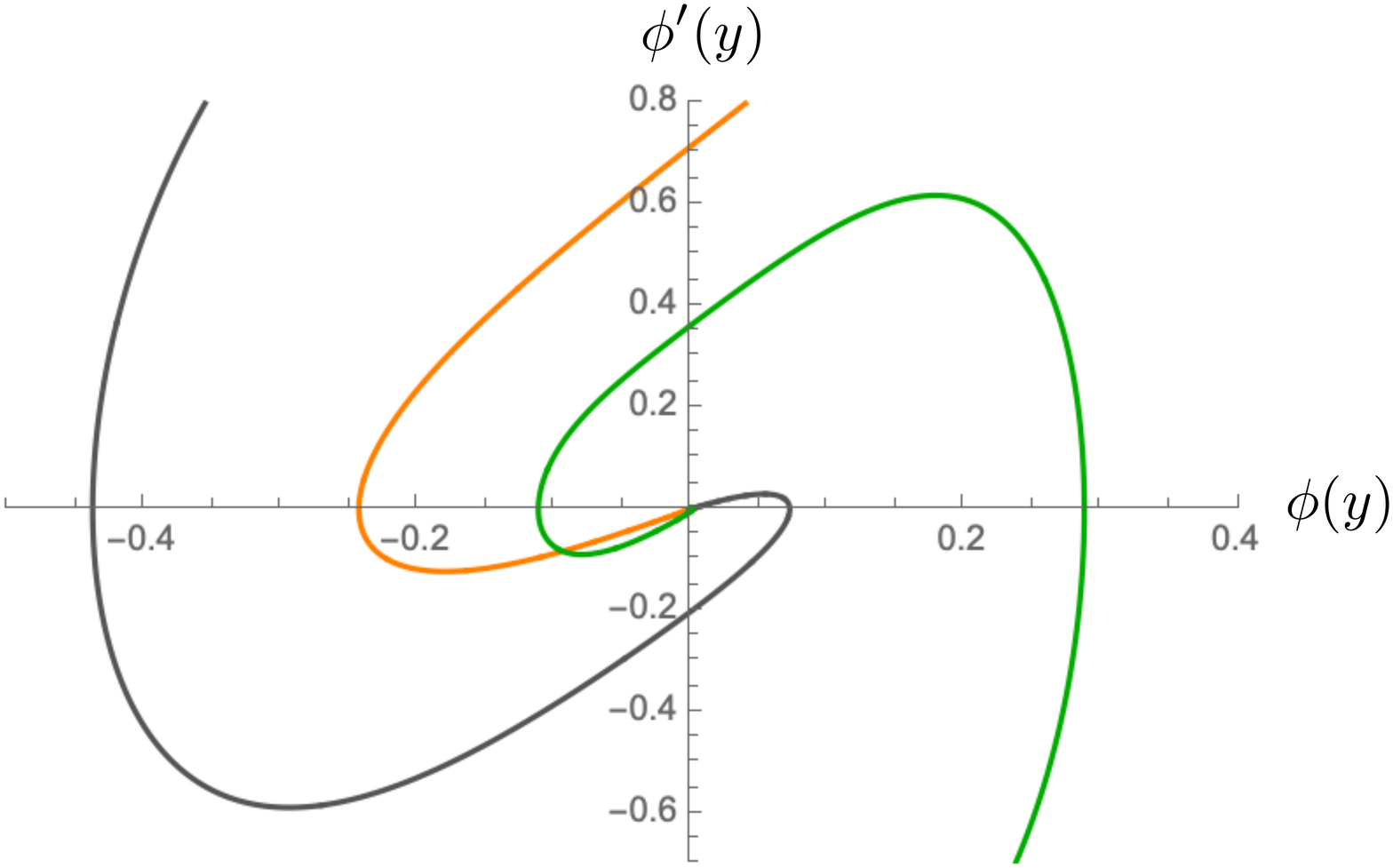}
 		\caption{}
 		\label{fig:phaseN2}
 	\end{subfigure}
 	\caption{Phase diagrams with respect to various negative values of $a$. Each curve has different number of critical field point $\phi_B$ (intersection point, $\phi\neq 0$ and $\phi'=0$) : one ($a=-3$), two ($a=-10$) and three ($a=-50$). From a numerical analysis, we find some regularity in which the number of the critical point increases when growing the negative value of $a$. The phase curve starts at the UV fixed point (the origin $\phi=0$) and, after clockwise flowing it stops at the IR fixed value of $\phi$.}
 	\label{fig:phaseN3}
 \end{figure}

So far, we discussed several different types of the RG flow relying on the value of $a$. Especially, the staircase RG flow appears as shown in Fig. \ref{fig:SolProfiles} for  $a=-0.55$. On the other hand, we showed that the bouncing RG flows occur for $a=-20$ in Fig. \ref{fig:SolProfiles} and for $a= - 10$ in Fig. \ref{fig:phaseB}. Now, we ask how many bouncing points exist in the bouncing RG flow. Although this question is very interesting, unfortunately answering this question looks very difficult because finding the number of bouncing points requires highly nontrivial nonperturbative analysis. We leave this issue to a future work, In this work, instead, we discuss the qualitative relation between the bouncing number and the intrinsic parameter $a$ by using the numerical analysis. In Fig. \ref{fig:phaseN3}, we depict the numerical behaviors of $\phi$ and $\phi'$ relying on the several other values of $a$ in the parameter regions of the bouncing flow. For the bouncing RG flow in Fig. \ref{fig:phaseN3}, a starting point ($\phi \ne 0$ and $\phi'=0$) and an ending point ($\phi = 0$ and $\phi'=0$) correspond to the UV and IR fixed points, respectively. Except for these two conformal fixed points, Fig. \ref{fig:phaseN3} shows that there exist more additional points satisfying $\phi'=0$ in the course of the RG flow. Those additional points exactly correspond to the bouncing points of the bouncing RG flow. Intriguingly, the numerical result in Fig. \ref{fig:phaseN3} indicates that the number of the bouncing points increases when the absolute value of the parameter $a$ increases in the bouncing RG flow region.

Before closing this section, there are some remarks on the relation between the bouncing RG flow and the cascading RG flow, which is another possible RG flow solution \cite{Kiritsis:2016kog}. In general, the cascading RG flow shows a very similar flowing behavior to the bouncing RG flow, though the cascading RG flow usually has an infinite number of the bouncing points. In spite of the similar flowing behavior, the cascading and bouncing RG flows have a big difference, which is the reason why a cascading RG flow does not appear in the present model. To understand this important difference between two RG flows, let us discuss the cascading RG flow more. For the cascading RG flow, it has been well known that there is no well-defined AdS space at a UV energy scale because the cascading RG flow does not allow a unitary CFT at a UV fixed point. On the dual gravity side, the scalar field representing the cascading RG flow usually violates the BF bound and then exhibits an oscillating behavior (tachyonic instabilities \cite{Weinberg:2012cd,Kiritsis:2016kog}) in the asymptotic region
\bea
\Phi(y)\simeq \alpha\,e^{dy/2}\cos\biggr(\frac{\vert \nu\vert}{2}y+\gamma\biggr),\,\,\,\,y\rightarrow -\infty,
\eea
where $\nu=\sqrt{4m^2R_{\text{uv}}^2+d^2}$ and $\alpha$ and $\gamma$ are integration constants. This is the typical feature usually appearing in an irrelevant deformation. The similar phenomenon without a bouncing behavior was studied in the ${\cal{N}}=1$ supergravity in type IIB string theory \cite{Klebanov:2000hb}. For the bouncing RG flow described by a relevant deformation, the corresponding scalar field is rapidly suppressed in the asymptotic region, so that its gravitational backreaction is usually negligible. However, the scalar field corresponding to the cascading RG flow does not suppressed in the asymptotic region, so that its gravitational backreaction inevitably modifies the asymptotic geometry. Therefore, the asymptotic geometry appearing in the cascading RG flow is not an AdS space. In other words, the cascading RG flow is UV incomplete similar to the $T\bar{T}$ deformation of a two-dimensional IR CFT \cite{Smirnov:2016lqw,Caselle:2013dra,McGough:2016lol,Park:2018snf}. On the other hand, since the bouncing RG flow is described by a relevant deformation, the dual field theory usually has a well defined CFT at a UV fixed point.


\newpage

\section{RG flow of the entanglement entropy} \label{sec5}

The $c$-theorem, as mentioned before, claims that the $c$-function representing the degrees of freedom of a system monotonically decreases along the RG flow. More accurately, there are three distinguished versions of the $c$-theorem conjecture \cite{Gukov:2015qea}. (1) The weakest version concerns the degrees of freedom only at the two endpoints of the RG flow such that $c_{\text{uv}}>c_{\text{ir}}$. (2) A stronger version asserts that $c$ is a monotonically decreasing function along the entire RG flow. (3) The strongest one claims that the RG flow is a gradient flow of the $c$-function. The last one still remains to be proven. Now, we focus on the first and second versions. 

In the previous sections, we discussed several different types of the RG flow which allows a nontrivial $\beta$-function. For the bouncing RG flow, in particular, the sing of the $\b$-function can have both  positive and negative values. This fact implies that the interaction strength of the dual field theory repeats increasing and decreasing successively. In this case, it would be interesting to ask how the $c$-function is affected by the change of the interaction strength and how the $c$-function evolves along the RG flow. In this section, we investigate the change of the $c$-function by using the holographic entanglement entropy.



Except for the free theories with a small perturbation, in general, it is a very difficult task to calculate  the entanglement entropy of an interacting field theory analytically. Even in this case, the holographic technique based on the AdS/CFT correspondence provide a very prominent tool which is useful to understand a non-perturbative features of strongly interacting systems. According to the Ryu-Takayanagi (RT) proposition \cite{Ryu:2006bv}, the entanglement entropy of the dual field theory has a one-to-one map to the area of the minimal surface extended to the bulk geometry. Now, we investigate the evolution of the entanglement entropy along the RG flow by using the RT formula.
To do so, we assume that the entangling points are located at $x=\pm l/2$. Then, a system is divided into a subsystem with $-l/2 \le x \le l/2$ and its complement. In this case, if the dual geometry is described by \eqref{met:normalcoord}, the area of the minimal surface is determined by
\bea		\label{eq:orgHEE}
S_E = \frac{1}{4 G} \int_{-l/2}^{l/2}  dx \sqrt{e^{2 A(y)} + y'^2} ,
\eea 
where $y$ is given by a function of $x$. Due to the invariance of the action under $x\to-x$, the minimal surface must have a turning point denoted by $y_*$ where $y'$ vanishes. After solving the equation of motion, the subsystems size can be reexpressed in terms of the turning point
\bea			\label{eq1:subsystemsize}
l =2 \int^\infty_{y_\ast} dy \ \frac{e^{A_*}}{e^{ A }\sqrt{e^{2 A}-e^{2 A_\ast }}},
\eea
where $A_\ast$ is the value of A at $y=y_*$. In addition, the entanglement entropy can also be rewritten as an integral form with the turning point
\bea	 \label{srt}
S_{E}=\frac{1}{2 G}\int_{y_\ast}^{\e_{\text{uv}}} dy \ \frac{e^{ A }}{\sqrt{e^{2 A }-e^{2 A_* }}} ,
\eea
where we introduce an appropriate UV cutoff $\e_{\text{uv}}$ to regulate a UV divergence.

After performing the above two integrals and rewriting the entanglement entropy in terms of the subsystem size $l$, we finally obtain the entanglement entropy in the UV region
\bea
S_A=\frac{c}{3}\log \frac{l}{\e_{\text{uv}}}+\delta (l) , \label{cft2} 
\eea
where $c=3 R/2G_N$ is the central charge of a dual two-dimensional CFT and $\delta (l)$ is a function depending on the subsystem size. This is exactly the form expected from a two-dimensional CFT \cite{Calabrese:2004eu}. Here, the first term is crucially relying on the UV theory from the RG flow viewpoint. If we consider a relevant deformation of a UV CFT like the various RG flows studied in this work, the first term always happens universally because the relevant deformation does not affect the UV theory. However, the second term above is not universal but crucially depends on the deformation.

In order to understand the IR physics beyond the UV regime, a non-perturbative method is required. If we are interested in the degrees of freedom of the IR physics, it can be represented as a holographic $c$-function based on the gauge/gravity duality. Using the holographic entanglement entropy, for two-dimensional QFT the $c$-function is defined by \cite{Casini:2004bw,Casini:2006es}
\bea
c =3\frac{dS_E(l)}{d\log l}=3\,l\frac{dS_E(l)}{d\,l}, \label{EEcfunction}
\eea
where $S_E(l)$ denotes the entanglement entropy evaluated with the subsystem size $l$. In this case the subsystem size is reinterpreted as the inverse of the energy scale observing the system, so that the system's energy scale moves from the UV to IR region when the subsystem size increases. This is exactly the same feature of the RG flow not in the momentum but real configuration space. One important result we need to note is that the $c$-function reduces to the central charge of a CFT at a conformal fixed point.

\begin{figure}
\centering
\includegraphics[width=0.57\textwidth]{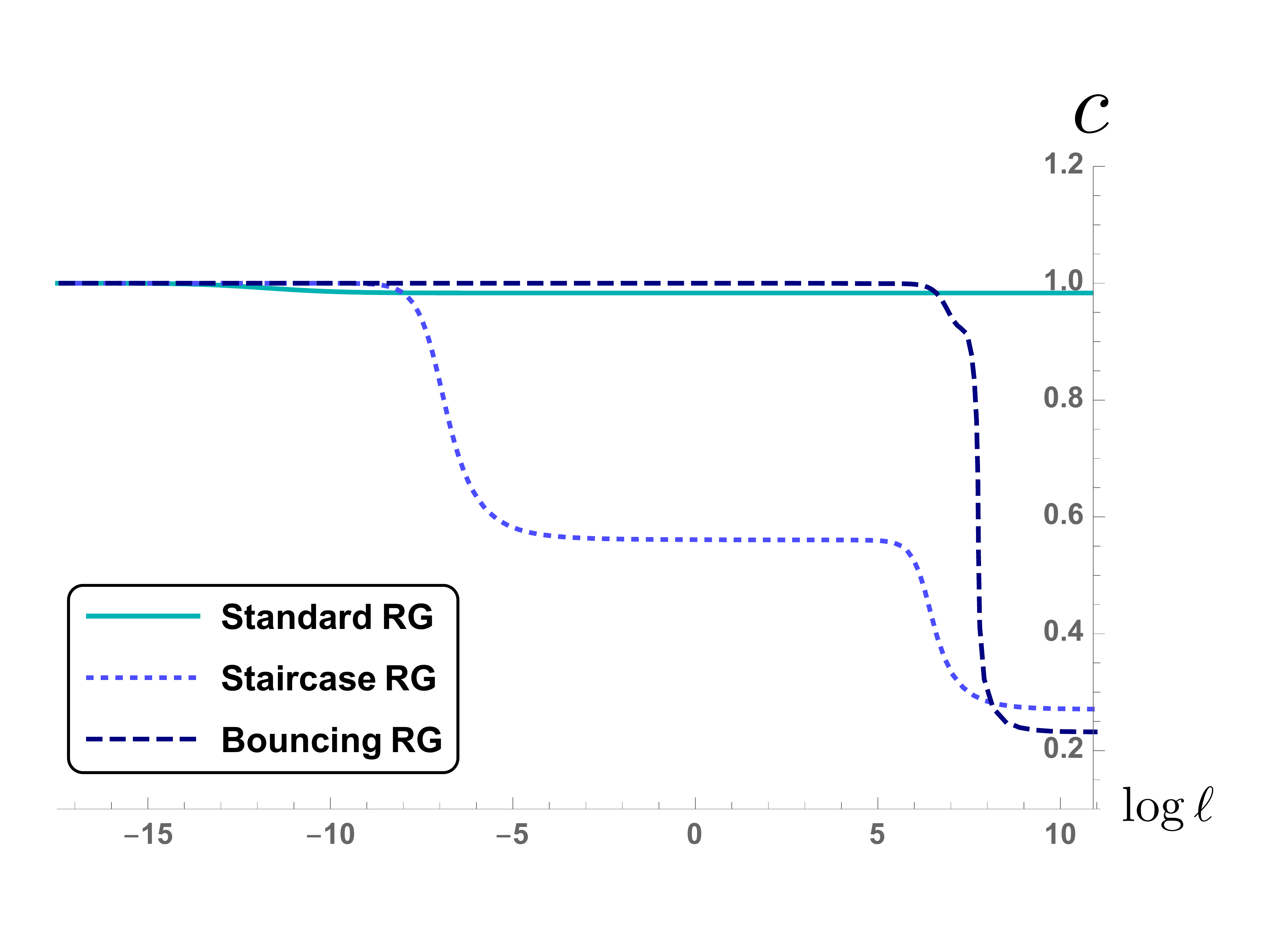}
\caption{The entropic $c$-theorems \eqref{cfunc} evaluated int the $\text{AdS}^{\text{uv}}_3\Rightarrow\text{AdS}^{\text{ir}}_3$ RG flows with different values of $a$. The reference central charge at the UV fixed point, $c_{\text{uv}}=1$. The values of $c_{\text{ir}}$ are $0.9833$ (standard RG), $0.2708$ (staircase RG) and $0.2319$ (bouncing RG), respectively.}
\label{fig:cfunc}
\end{figure}

In a deformed $\text{AdS}_3$ space, in general, the $c$-function can be formally rewritten as \cite{Myers:2012ed}
\bea
c=\frac{3\, dS_E}{d\log l}=\frac{3}{4G_N}\frac{l}{\gamma(l)},\label{cfunc}
\eea
where $\gamma(l)$ is related to a conserved quantity appearing in \eqref{eq:orgHEE}. Since the holographic entanglement entropy in \eqref{eq:orgHEE} does not explicitly depend on $x$, there is one conserved quantity which at the turning point is given by
\bea
\gamma(y_\ast)=e^{-A_\ast}.
\eea 
In this case, since the position of the turning point crucially relies on the subsystem size, $\gamma(y_\ast)$ can be represented as a function of $l$ instead of $y_\ast$ by using \eqref{eq1:subsystemsize}. In Fig. \ref{fig:cfunc}, we plot several exact $c$-functions appearing in the standard, staircase and bouncing RG flows with different values of $a$. The result shows that the $c$-function always monotonically decreases independent of the type of the RG flow. For the bouncing RG flow with both positive and negative $\b$-function, the numerical result shows that the $c$-function always decreases monotonically along the RG flow regardless of the strength of the interaction. 

\begin{figure}
\centering
\begin{subfigure}{.47\textwidth}
  \centering
  \includegraphics[width=1\linewidth]{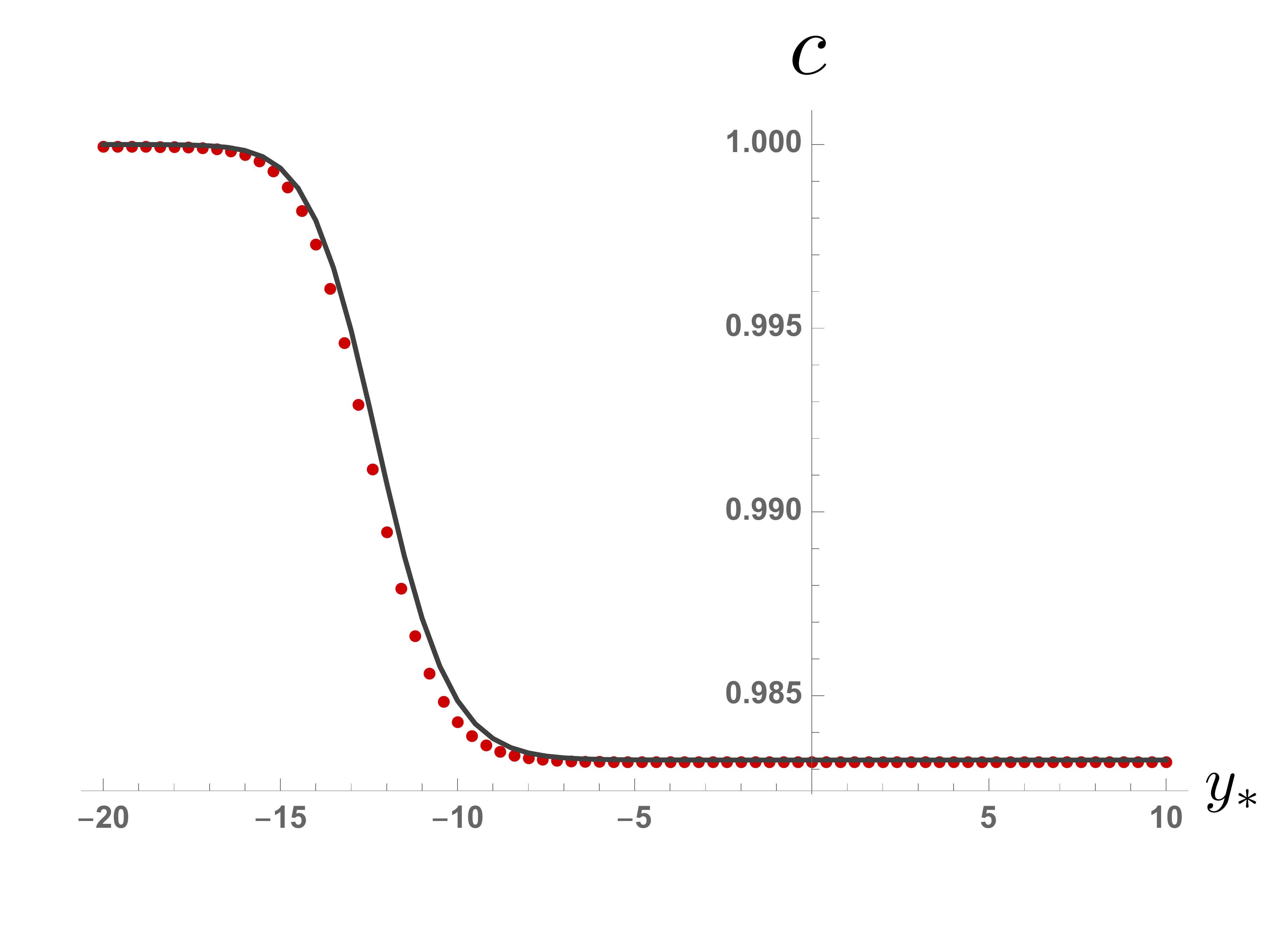}
  \caption{}
  \label{fig:NumP1}
\end{subfigure}%
\begin{subfigure}{.47\textwidth}
  \centering
  \includegraphics[width=1\linewidth]{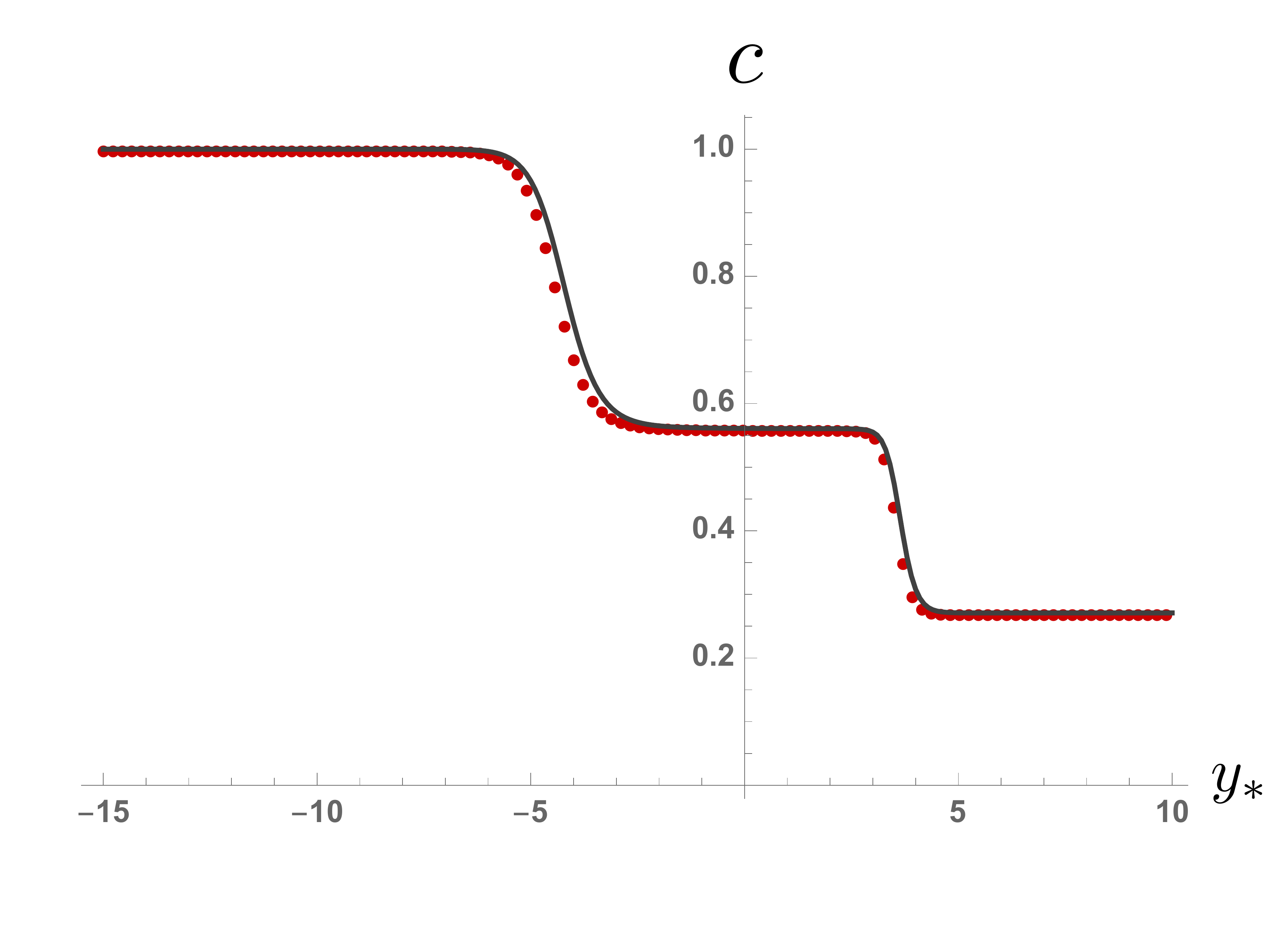}
  \caption{}
  \label{fig:NumP1}
\end{subfigure}
\begin{subfigure}{.47\textwidth}
  \centering
  \includegraphics[width=1\linewidth]{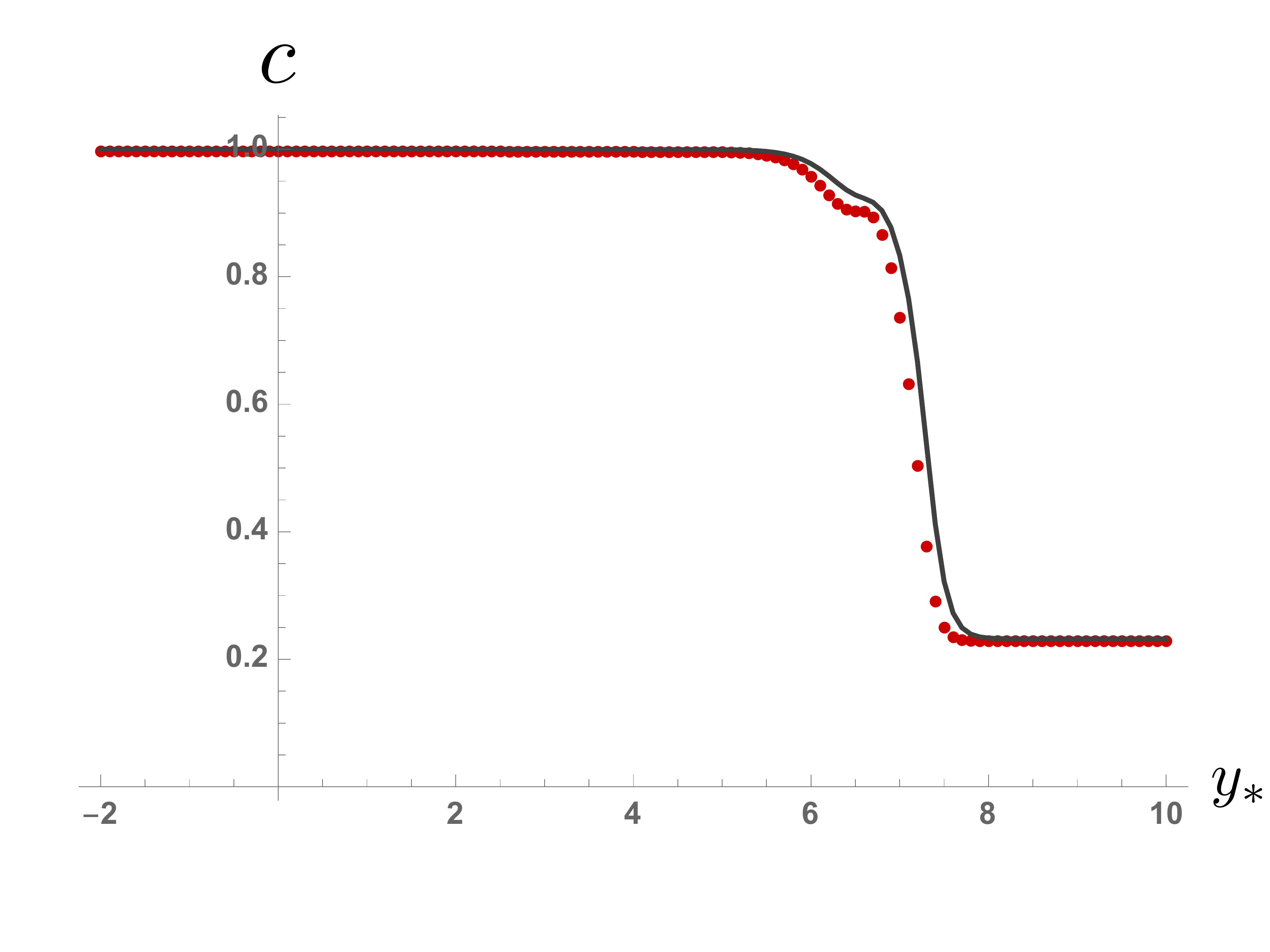}
  \caption{}
  \label{fig:NumP1}
\end{subfigure}
\caption{Comparing holographic $c$-functions : (a) standard RG, (b) staircase RG and (c) bouncing RG, where the red dotted line represents the result of \eqref{hcfunction} and the black line represents the entanglement $c$-function \eqref{cfunc}. Here we fix the central charge to be $1$ at the UV fixed point.} 
\label{fig:NumPlots}
\end{figure}

For consistency check, in Fig. \ref{fig:NumPlots} we also compare the $c$-function in \eqref{cfunc}, which was derived from the holographic entanglement entropy, with \eqref{hcfunction} obtained in the holographic renormalization procedure. The result in Fig. \ref{fig:NumPlots} shows that two $c$-functions defined by two different ways have a small discrepancy in the intermediate energy scale. Nevertheless, these two $c$-functions
shows the qualitatively same flowing behavior in the entire energy scale and reduces to the exact same central charges of the UV and IR CFTs at the fixed points. Although two $c$-functions defined here have a small discrepancy in quantity, they are qualitatively almost equivalent and represents the RG flow expected by the $c$-theorem.

\section{Discussion} \label{discussion}

We investigated several different types of the RG flow which can appear in a two-dimensional deformed CFT. In order to realize such RG flows holographically, we took into account a three-dimensional dual gravity theory with a specific scalar potential which allows many local maxima and extrema. Due to the invariance of the potential under $\phi \to - \phi$, the potential usually allows a local extremum at $\phi = 0$ which was identified with the UV fixed point of the dual field theory. Near this UV fixed point, if we restrict for the scalar field to have a mass only in the range of $-1< m^2 R_{\text{uv}}^2 <0$, the local extremum at $\phi=0$ becomes a local maximum or a unstable equilibrium point and, on the dual field theory side, the corresponding dual scalar operator becomes a relevant operator. Although the effect of a relevant operator is negligible in the UV regime, it causes a nontrivial RG flow and seriously modifies the IR physics. On the dual gravity side, the RG flow caused by a relevant scalar operator can be matched to the rolling of the bulk scalar field. Since the potential considered here allows a local minimum near the local maximum defined at $\phi=0$ except only for $a=0$, the scalar field naturally rolls down to a local minimum which corresponds to a new IR fixed point of the deformed dual field theory. As a consequence, the rolling of the scalar field from the unstable to new stable equilibrium points describes the RG flow of the dual field theory from the UV to IR fixed points. From the field theory point of view, since this RG flow is highly nonperturbative, it is usually a very difficult task to understand the details of the RG flow in the entire region of the energy scale. However, there is still a chance to investigate the nonperturbative feature of the RG flow by using the holographic dual of this RG flow. In the present work, we have studied the possible RG flows of the quantum field theory deformed by a relevant scalar operator. Interestingly, we showed that the toy model we considered leads to several different types of the RG flow relying on the value of the intrinsic parameter $a$. The resulting RG flows can be summarized as follows:

\begin{itemize}

\item For $a>0$, the RG flow of the dual field theory is described by the standard RG flow, in which the $\b$-function is always negative and does not oscillate at the intermediate energy scale. From this result, we can see that the coupling constant of the dual field theory increases monotonically along the RG flow, while the $c$-function corresponding to the degrees of freedom monotonically decreases. 

\item For $a=0$, we showed that there is no local minimum near the local maximum corresponding the UV fixed point. This implies that the RG flow is not terminated due to the absence of the IR fixed point. Therefore, the RG flow for $a=0$ becomes an IR incomplete RG flow.

\item  For $a<0$, a new IR fixed point appears again, so that the corresponding RG flow is IR complete. In this parameter region, the resulting RG flow shows two different flowing behaviors from the previous standard RG flow. When the absolute value of $a$ is very small, the staircase RG flow appears. The $\b$-function of the staircase RG flow is always negative except for two fixed points, which is similar to the previous standard RG flow. Unlike the standard RG flow, however, the $\b$-function of the staircase RG flow oscillates. Therefore, the coupling constant of the staircase RG flow increases monotonically along the RG flow but repeats the fast and slow increasing due to the oscillation of the $\b$-function. For a large absolute value of $a$, the bouncing RG flow occurs. In general, the bouncing RG flow has two branching solutions. One has a positive $\b$-function, while the other branch has a negative value. At the bouncing points, the RG flow changes the branch with changing the sign of the $\b$-function. On the dual field theory side, this feature shows that the coupling constant increasing in one branch becomes decreasing in the other branch after passing through the bouncing point. As a result, the interaction strength of the bouncing RG flow does not increases monotonically along the RG flow. We finally showed that the number of the bouncing points in the bouncing RG flow increases as the absolute value of $a$ increases.

\end{itemize}

We also studied the $c$-function relying on the energy scale by using the holographic entanglement entropy technique. A variety of the RG flows we found have a nontrivial $\b$-function which determines the strength of the coupling constant. For the bouncing RG flow, since the $\b$-function has a positive value at an intermediate energy scale, the interaction strength can decrease in the course of the RG flow. At this energy scale, it would be interesting to ask how the $c$-function behaves. By applying the holographic entanglement entropy technique, in the present work, we studied the change of the $c$-function of various RG flows we found. Intriguingly, we numerically showed  that the $c$-function always decreases along the RG flow regardless of the type of the RG flow. Even in the bouncing RG flow which allows the decreasing coupling constant, for example, the $c$-function monotonically decreases with satisfying the $c$-theorem. In the present work, we focused on only the two-dimensional deformed CFT.  However, it would be more interesting to know the possible RG flows of a higher dimensional deformed CFT by applying the methods used in this work. We hope to report more interesting results in future works.

\bigskip

\section*{Acknowledgments}
We are grateful to Yunseok Seo for insightful discussions.
This work was supported by Basic Science Research Program through NRF grant No. NRF-2016R1D1A1B03932371 and by Mid-career Researcher Program through the National Research Foundation of Korea grant No. NRF-2019R1A2C1006639. 
J. Lee was supported by Basic Science Research Program through the National Research Foundation of Korea funded by the Ministry of Education grant No. NRF-2018R1A6A3A11049655.






	\bibliographystyle{apsrev4-1}

\bibliography{ERG8}



\end{document}


\bibitem{Smirnov:2016lqw} 
F.~A.~Smirnov and A.~B.~Zamolodchikov,
Nucl.\ Phys.\ B {\bf 915}, 363 (2017)
doi:10.1016/j.nuclphysb.2016.12.014
[arXiv:1608.05499 [hep-th]].

\bibitem{Caselle:2013dra} 
M.~Caselle, D.~Fioravanti, F.~Gliozzi and R.~Tateo,
JHEP {\bf 1307}, 071 (2013)
doi:10.1007/JHEP07(2013)071
[arXiv:1305.1278 [hep-th]].

\bibitem{McGough:2016lol} 
L.~McGough, M.~Mezei and H.~Verlinde,
JHEP {\bf 1804}, 010 (2018)
doi:10.1007/JHEP04(2018)010
[arXiv:1611.03470 [hep-th]].

\bibitem{Park:2018snf} 
C.~Park,
Int.\ J.\ Mod.\ Phys.\ A {\bf 33}, no. 36, 1850226 (2019)
doi:10.1142/S0217751X18502263
[arXiv:1812.00545 [hep-th]].
\\